\documentclass[useAMS,usenatbib]{mn2e}
\usepackage{epsfig}
\usepackage{graphicx}
\usepackage{epstopdf}
\DeclareGraphicsExtensions{.ps}
\usepackage{times}
\usepackage{natbib}
\usepackage{amssymb}
\usepackage{amsmath}
\usepackage{tabularx}
\usepackage{amsfonts,amsmath,amssymb}
\usepackage{txfonts}
\usepackage{graphicx}
\usepackage{color}
\usepackage{flushend}
\usepackage{enumitem}
\usepackage{tabularx}
\usepackage{color}
\usepackage{longtable}
\usepackage{subfigure}
\usepackage{supertabular}
\usepackage[font=small,labelfont=bf,tableposition=top]{caption}
\usepackage{booktabs,threeparttable}
\usepackage{placeins}
\usepackage{float}
\floatstyle{plaintop}
\restylefloat{table}
\usepackage{afterpage}
\usepackage[colorlinks=true,linkcolor=red, citecolor= blue, urlcolor=cyan]{hyperref} 
\newif\ifAMStwofonts
\AMStwofontstrue
\newcommand {\aplt} {\ {\raise-.5ex\hbox{$\buildrel<\over\sim$}}\ } 
\newcommand{\be}{\begin{equation}}
\newcommand{\ee}{\end{equation}}

\newcommand{\mincir}{\raise
  -2.truept\hbox{\rlap{\hbox{$\sim$}}\raise5.truept \hbox{$<$}\ }}
\newcommand{\magcir}{\raise
  -2.truept\hbox{\rlap{\hbox{$\sim$}}\raise5.truept \hbox{$>$}\ }}
\newcommand{\siml}{\raise
  -2.truept\hbox{\rlap{\hbox{$\sim$}}\raise5.truept \hbox{$<$}\ }}
\newcommand{\simg}{\raise
  -2.truept\hbox{\rlap{\hbox{$\sim$}}\raise5.truept \hbox{$>$}\ }}
\newcommand{\Mszsigma}{$M_{200}^\mathrm{SZ+\sigma}$}
\newcommand{\MszPlanck}{$M_{200}^\mathrm{SZ+Planck}$}
\newcommand{\Mdyn}{$M_{200}^\mathrm{dyn}$}
\newcommand{\Rdyn}{$R_{200}^\mathrm{dyn}$}

\title[Galaxy Kinematics and Masses of Clusters to z=1.3]{Galaxy Kinematics and Mass Calibration in Massive SZE Selected Galaxy Clusters to z=1.3}

\newcommand{\Munich}{$^{1}$}
\newcommand{\ExcellenceCluster}{$^{2}$}
\newcommand{\Trieste}{$^{3}$}
\newcommand{\MPE}{$^{4}$}
\newcommand{\KICPChicago}{$^{5}$} 
\newcommand{\ANL}{$^{6}$} 
\newcommand{\MIT}{$^{7}$} 
\newcommand{\AAUChicago}{$^{8}$}
\newcommand{\FNAL}{$^{9}$}
\newcommand{\PhysicsUChicago}{$^{10}$}
\newcommand{\Miss}{$^{11}$}
\newcommand{\EFIChicago}{$^{12}$}
\newcommand{\Taipei}{$^{13}$} 
\newcommand{\Berkeley}{$^{14}$} 
\newcommand{\McGill}{$^{15}$} 
\newcommand{\StanfordKPAC}{$^{16}$}
\newcommand{\StanfordPhys}{$^{17}$}
\newcommand{\UMon}{$^{18}$}
\newcommand{\StonyBrook}{$^{19}$} 
\newcommand{\DARK}{$^{20}$} 
\newcommand{\Colorado}{$^{21}$} 
\newcommand{\NASA}{$^{22}$} 
\newcommand{\UM}{$^{23}$} 
\newcommand{\Michigan}{$^{24}$}
\newcommand{\CfA}{$^{25}$}
\newcommand{\Davis}{$^{26}$}
\newcommand{\LLNL}{$^{27}$}
\newcommand{\Chile}{$^{28}$}

\author[Capasso et al.] {R.~Capasso\thanks{Raffaella.Capasso@physik.lmu.de}\Munich$^,$\ExcellenceCluster,
A.~Saro\Munich$^,$\ExcellenceCluster$^,$\Trieste,
J.~J.~Mohr\Munich$^,$\ExcellenceCluster$^,$\MPE,
A.~Biviano\Trieste,
S.~Bocquet\KICPChicago$^,$\ANL, 
V.~Strazzullo\Munich, 
\newauthor
S.~Grandis\Munich$^,$\ExcellenceCluster, 
D.~E.~Applegate\KICPChicago,
M.~B.~Bayliss\MIT,
B.~A.~Benson\KICPChicago$^,$\AAUChicago$^,$\FNAL,
L.~E.~Bleem\KICPChicago$^,$\ANL$^,$\PhysicsUChicago,
\newauthor
M.~Brodwin\Miss,
E.~Bulbul\MIT,
J.~E.~Carlstrom\KICPChicago$^,$\ANL$^,$\AAUChicago$^,$\PhysicsUChicago$^,$\EFIChicago,
I.~Chiu\Taipei,
J.~P.~Dietrich\Munich$^,$\ExcellenceCluster,
N.~Gupta\Munich$^,$\ExcellenceCluster,
\newauthor
T.~de~Haan\Berkeley$^,$\McGill,
J.~Hlavacek-Larrondo\StanfordKPAC$^,$\StanfordPhys$^,$\UMon,
M.~Klein\Munich$^,$\MPE,
A.~von~der~Linden\StanfordKPAC$^,$\StanfordPhys$^,$\StonyBrook$^,$\DARK,
\newauthor
M.~McDonald\MIT,
D.~Rapetti\Munich$^,$\ExcellenceCluster$^,$\Colorado$^,$\NASA,
C.~L.~Reichardt\UM,
K.~Sharon\Michigan,
B.~Stalder\CfA,
\newauthor
S.~A.~Stanford\Davis$^,$\LLNL,
A.~A.~Stark\CfA,
C.~Stern\Munich$^,$\ExcellenceCluster,
A.~Zenteno\Chile
\\
\\
\date{\textit{Affilitations are listed at the end of the paper}}
}

\begin{document}
\date{\today}

\maketitle

\label{firstpage}      

\begin{abstract} 
The galaxy phase-space distribution in galaxy clusters provides insights into the formation and evolution of cluster galaxies, and it can also be used to measure cluster mass profiles.  We present a dynamical study based on $\sim$3000 passive, non-emission line cluster galaxies drawn from 110 galaxy clusters. The galaxy clusters were selected using the Sunyaev-Zel'dovich effect (SZE) in the 2500~deg$^2$ SPT-SZ survey and cover the redshift range $0.2 < z < 1.3$.
We model the clusters using the Jeans equation, while adopting NFW mass profiles and a broad range of velocity dispersion anisotropy profiles.  The data prefer velocity dispersion anisotropy profiles that are approximately isotropic near the center and increasingly radial toward the cluster virial radius, and this is true for all redshifts and masses we study.  The pseudo-phase-space density profile of the passive galaxies is consistent with expectations for dark matter particles and subhalos from cosmological $N$-body simulations.  The dynamical mass constraints are in good agreement with external mass estimates of the SPT cluster sample from either weak lensing, velocity dispersions, or X-ray $Y_X$ measurements. 
However, the dynamical masses are lower (at the 2.2$\sigma$ level) when compared to the mass calibration favored when fitting the SPT cluster data to a $\Lambda$CDM model with external cosmological priors, including CMB anisotropy data from Planck. The discrepancy grows with redshift, where in the highest redshift bin the ratio of dynamical to SPT+Planck masses is $\eta=0.63^{+0.13}_{-0.08}\pm0.06$ (statistical and systematic), corresponding to a $2.6\sigma$ discrepancy.
\end{abstract}



\section{Introduction}
\label{sec:introduction}

In the current paradigm of structure formation, halos form through the gravitational collapse of overdense regions that are seeded by processes in the early universe. The formation of cold dark matter (CDM) dominated halos proceeds through a sequence of mergers and the accretion of surrounding material, leading to the formation of the galaxy groups and clusters we observe.  Baryonic processes associated with the intracluster medium (ICM) and the galaxies also play a role, making galaxy clusters important laboratories for investigations of structure formation and galaxy evolution as well as useful cosmological probes.

Studies of structure formation using cosmological $N$-body simulations have been used to demonstrate that halos formed from collisionless CDM have, on average, a universal mass density profile \citep[][hereinafter NFW]{NA96.1, NA97.1}. This profile is characterized by two parameters: the virial radius $r_{200}$\footnote{$r_{\Delta}$ defines the sphere within which the cluster overdensity with respect to the critical density at the cluster redshift is $\Delta$. Throughout this paper, we consider $\Delta = 200$ and refer to $r_{200}$ simply as the virial radius.}, and the scale radius $r_\mathrm{s}$, which is the radius at which the logarithmic slope of the density profile is $-2$.  Numerous observational studies have found the mass distributions of clusters to be well described by this model \citep[e.g.,][]{1997Carlberg, vanderMarel2000, 2003Biviano, Katgert2004, 2014Umetsu}. 

Another interesting feature is the finding in $N$-body simulations that the quantity $ \rho / \sigma^{3}$, where $\rho$ is the mass density and $\sigma$ the velocity dispersion, has a power-law form. This quantity is known as pseudo-phase-space density (PPSD) profile, $Q(r)$, and its power-law form resembles that of the self-similar solution for halo collapse by \citet{1985Bertschinger} and is thought of as a dynamical equivalent of the NFW mass density profile \citep{2001Taylor}. 
Others \citep{2005Austin, 2006Barnes} have suggested that the PPSD profile results from dynamical collapse processes, and should therefore be a robust feature of approximately virialized halos that have undergone violent relaxation \citep{1967LyndenBell}.  

The galaxy population is more difficult to study in simulations, because of the over-merging problem, i.e. the premature destruction of dark matter halos in the dense clusters environments in dissipationless $N$-body simulations \citep[e.g.][]{1996Moore}, and the additional baryonic physics that must be included.  However, from the observational side the properties of the galaxy population and trends with mass and redshift can be readily measured and interpreted as long as:  (1) selection effects are understood and (2) precise cluster mass measurements are available to ensure that the same portion of the virial region is being studied in clusters of all masses and redshifts.  By comparing the galaxy properties to the expectations for collisionless particles studied through $N$-body simulations, one can characterize the impact of possible additional interactions beyond gravity that are playing a role in the formation of the galaxy population.

As an example, the radial distribution of galaxies in clusters is well fit by an NFW model when clusters are stacked in the space of $r/r_\Delta$ \citep[e.g.][]{lin04a, 2007Muzzin, 2014vanderBurg,2015vanderBurg,2016Zenteno}.  In cluster samples extending to redshift $z\sim1$, it is clear that the concentration $c_\Delta$, defined as the ratio between $r_\Delta$ and $r_{s}$, varies dramatically from cluster to cluster, and that when stacked, the $c_\Delta$ varies systematically with the prevalence of star formation \citep[][hereafter H17]{2017Hennig}.  The red, passively evolving galaxies have concentrations similar to those expected for the dark matter on these halo mass scales, while the star forming, and presumably infalling blue galaxies are far less concentrated. The number of luminous cluster galaxies (magnitudes $m<m_*+2$) within the virial region scales with cluster mass as $N\propto M^\alpha$ where $\alpha\sim0.85$ \citep{lin04a}, and this property appears to be unchanged since redshift $z\sim1$ (H17).  The departure from $\alpha=1$ in this relation is puzzling, given that massive clusters accrete lower mass clusters and groups \citep{lin04b} and this is presumably evidence for galaxy destruction processes that are more efficient in the most massive halos or for a mass accretion history that varies with mass on cluster scales \citep[see discussion in][]{2016Chiu-b}.  There is evidence for an increase in the fraction of cluster galaxies that are dominated by passively evolving stellar populations since $z\sim1$ (H17), and this observed increase provides constraints on the timescales over which quenching of star formation occurs in those galaxies that are accreted by clusters \citep[see, e.g.,][]{mcgee09}. 

Understanding the dynamics of galaxy accretion into clusters, from either lower mass clusters and groups or even individual systems from within the surrounding low density region, can shed additional light on galaxy evolution.  A simulation based study argues that satellite orbits should become marginally more radial at higher redshifts, especially for systems with a higher host halo mass \citep{2011Wetzel}.  Probes of redshift trends in the orbital characteristics of cluster galaxies have already been carried out \citep{2009Biviano}, providing some indication that passive galaxies have systematically different orbits at low and high redshift. In other studies of high redshift, relatively low mass systems, evidence has emerged that recently quenched galaxies have a preferred phase space distribution that is different from that of passive galaxies \citep{muzzin14, 2016Noble}.

In this paper, we attempt to build upon these studies by focusing on a dynamical analysis of galaxies within a large ensemble of Sunyaev-Zel'dovich effect (SZE) selected galaxy clusters extending to redshift $z\sim1.3$. In contrast to these previous dynamical studies, our cluster sample has a well understood selection that does not depend on the galaxy properties, and the sample extends over a broad redshift range, allowing a cleaner examination of redshift trends.  Moreover, each cluster has an SZE based mass estimate with $\sim$25~percent uncertainty \citep{bocquet15}, enabling us to estimate virial radii $r_\Delta$ with $\sim8$~percent uncertainties and thereby ensuring that we are examining comparable regions of the cluster at all masses and redshifts.  

Our goals are to study (1) whether there is evidence that the orbital characteristics of the passive galaxies are changing with redshift or mass in the cluster ensemble, (2) whether there is evidence within the galaxy dynamics for dynamical equilibrium and self-similarity with mass and redshift, and (3) whether the cluster mass constraints from our analysis are consistent with masses obtained through independent calibration in previously published SPT analyses.   

Combining spectroscopic observations obtained at Gemini South, the VLT and the Magellan telescopes in a sample of 110 SPT-detected galaxy clusters, we construct a large sample of $\sim$3000 passive cluster members, spanning the wide redshift range of $0.2  < z < 1.3$.  With this dataset we carry out a Jeans analysis \citep[e.g.][]{1987Binney} that adopts a framework of spherical symmetry and allows for a range of different velocity dispersion anisotropy profiles.  Specifically, we use the Modeling Anisotropy and Mass Profiles of Observed Spherical Systems code \citep[][hereafter MAMPOSSt]{2013MAMPOSSt} to explore the range of models consistent with the data, and then we use the results to characterize the velocity dispersion anisotropy profile, to test for evidence of virialization with the pseudo-phase-space density (PPSD) profile and to probe for trends with cluster mass or redshift in both.  Exploring a broad range of possible velocity dispersion anisotropy profiles then allows us to extract robust constraints on the cluster virial masses as well.  Throughout this paper, we address a number of limitations that have to be taken into account, such as the degeneracy between the mass and the velocity anisotropy profiles (see Section~\ref{sec:theory}), the assumptions of spherical symmetry and dynamical equilibrium, and the presence of foreground/background interloper galaxies projected onto the cluster virial region. \citet{2013MAMPOSSt} have tested the accuracy of MAMPOSSt by analysing a sample of clusters extracted from numerical simulations, recovering $r_{200}$ estimates with mean bias at $\le 2.5$\% and {\it rms} scatter of 6\% for kinematic samples with 500 tracers.  

The paper is organized as follows: In Section~\ref{sec:theory} we give an
overview of the theoretical framework. In Section~\ref{sec:data} we summarize the dataset used for our analysis. The results are presented in Section~\ref{sec:results}, where we discuss the outcome of our analysis of the velocity dispersion anisotropy profile, the PPSD profiles, the virial mass comparisons, and the impact of disturbed clusters on our analysis.  We present our conclusions in Section~\ref{sec:conclusions}.  Throughout this paper we adopt a flat $\Lambda$CDM cosmology with the Hubble constant $H_{0} = 70 \,  \text{km} \, \text{s}^{-1} \,  \text{Mpc}^{-1}$, and assume the matter density parameter $\Omega_{\text{M}} = 0.3$. The virial quantities are computed at radius $r_{200}$. All quoted uncertainties are equivalent to Gaussian $1\sigma$ 
confidence regions, unless otherwise stated.


\section{Theoretical Framework}
\label{sec:theory}

The dynamical analysis implemented in this paper is based on the application of the Jeans equation to spherical systems \citep{1987Binney}. 
In spherical coordinates, the Jeans equation can be written as
\begin{equation}                                                                
\label{eqn:jeans}                                                              
\frac{d(\nu \sigma_{r}^{2})} { d r } + \frac{\nu}{r} \left[ 2 { \sigma_{r}^{2}} - ({ \sigma_{\theta}^{2}} + {\sigma_{\phi}^{2}} ) \right]   = - \nu \frac{d \Phi}{dr}  ,
\end{equation}  
where $\nu$ is the number density profile of the tracer galaxy population, $\Phi$ is the gravitational potential, and $ { \sigma_{r}}, { \sigma_{\theta}}, { \sigma_{\phi}} $ are the components of the velocity dispersion along the three spherical coordinates $r, \theta, \phi$. 
It is convenient to write this equation as
\begin{equation}                                                                
\label{eqn:jeans2}                                                              
\frac{GM(r)}{r} = - {\sigma_{r}^{2}} \left( \frac{d \ln \nu}{ d \ln r } + \frac{d \ln {\sigma_{r}^{2}}} { d \ln r} + 2 \beta \right) ,
\end{equation} 
where $M(r)$ is the enclosed mass within radius $r$, $G$ is Newton's constant, $\beta \equiv 1 - (\sigma_{\theta}^{2} / \sigma_{r}^{2})$ is the velocity dispersion anisotropy that is generically a function of radius, and $\sigma_{\theta}= \sigma_{\phi}$.

In principle, it is therefore possible to use Eq.~\ref{eqn:jeans2} to estimate the mass distribution $M(r)$ of the system. However, we as external observers can measure only projected quantities, such as the surface density profile of the tracer population $\Sigma(R)$ and the line-of-sight (LOS) velocity distribution (or, alternatively, the line-of-sight velocity dispersion $\sigma_{\text{LOS}}(R)$).  These two observed functions are not sufficient to derive a unique mass model--- at least within the context of the typical observational uncertainties where full knowledge of the line of sight velocity distribution is lacking \citep{1987Merritt}. This degeneracy between the mass and the velocity anisotropy profiles can be addressed in several ways. 

\subsection{Dynamical analysis with MAMPOSSt}
\label{sec:MAMPOSSt}

A first method consists of assuming that, at a given projected radius, the LOS velocity distribution can be described by a Gaussian \citep{2008Strigari, 2010Wolf}. However, because the true distribution may deviate from Gaussianity--- including the case of anisotropic systems \citep{1987Merritt}--- the constraints from this approach on the anisotropy are weak \citep{2009Walker}. A step forward, therefore, consists of analysing the kurtosis of the LOS velocity distribution. This was found to be a powerful tool to break the mass-anisotropy degeneracy \citep{1987Merritt,1993Gerhard, 1993VdM, 1993Zabludoff, 2002Lokas, 2003Lokas}.
Much effort has been put into further constraining the anisotropy taking into account more of the information contained in the projected phase-space  velocity distribution as a function of projected radius, considering, for example,  the full set of even moments \citep{2000Kronawitter, 2008Wojtak, 2009Wojtak}.

In this work, we fit the whole projected phase-space velocity distribution using the MAMPOSSt code \citep{2013MAMPOSSt}.  We use this code to determine the mass and anisotropy profiles of a cluster in parametrized form by performing a likelihood exploration to the distribution of the cluster galaxies in projected phase-space, constraining the parameters describing these two profiles. 
This method is based on the assumption of spherical symmetry, and adopts a Gaussian as the shape of the distribution of the 3D velocities without demanding Gaussian LOS velocity distributions. We emphasize that it does not assume that light traces mass, allowing the scale radius $r_{\text{s}}$ of the total mass distribution to differ from that of the galaxy distribution. 
For more details on the code, we refer the reader to \citet{2013MAMPOSSt}.

MAMPOSSt requires parametrized models for the number density profile, the mass profile and the velocity anisotropy profile, without any limitation on the choice of these models.  We will address the issue of the number density profile in Section~\ref{sec:nR}.

The current estimates of the systematic error on MAMPOSSt derived dynamical masses are $\approx10\%$, where this number comes from an analysis of clusters extracted from numerical simulations \citep[see][]{2013MAMPOSSt}.  In their study, they find that for the dynamical tracers defined to lie within a sphere of $r_{100}$ that the estimate of the virial radius $r_{200}$ is biased at $\le3.3$\% \citep[see Table 2;][]{2013MAMPOSSt}.  Thus, as a prior on the virial mass $M_{200}$ bias, we adopt a Gaussian with $\sigma=10$\% centered at no bias.

\subsection{Mass and anisotropy profiles}
\label{sec:profiles}

For the mass profile in our analyses, 
we consider 5 models, namely the Navarro, Frenk and White profile \citep[NFW;][] {1996NFW}, the Einasto profile \citep{1965Einasto}, the Burkert profile \citep{1995Burkert}, the Hernquist profile \citep{1990Hernquist}, and the Softened Isothermal Sphere \citep[SIS;][]{1999Geller}.
All these models have been applied to galaxy clusters in previous works \citep[e.g.][]{mohr96,2003AJ....126.2152R, Katgert2004, 2006Rines, 2006A&A...456...23B}.  As we show later in Section~\ref{sec:results}, our data cannot distinguish among these different mass profiles, and so in the analyses described below we adopt the NFW model 
\be
\rho(r)=\rho_0 \left({r\over r_\mathrm{s}}\right)^{-1}\left(1+{r\over  r_\mathrm{s}}\right)^{-2}
\ee
where $\rho_0$ is the central density and $r_\mathrm{s}$ is the scale radius where the logarithmic derivative of the density profile reaches -2.

For the velocity anisotropy profile, we consider the following five models that have been used in previous MAMPOSSt analyses:
\begin{enumerate}[label=(\alph*),leftmargin=*,align=left] 
\item[C] a radially constant anisotropy model, 
\be
\beta_{C}(r)= \theta_{\beta};  
\ee
\item[T] the Tiret anisotropy profile \citep{2007Tiret}, 
\be 
\beta_{\text{T}}(r)=\theta_{\beta}\dfrac{r}{r+r_\mathrm{s}}, 
\ee
which is isotropic at the center and characterized by the anisotropy value $\theta_{\beta}$ at large radii. The transition radius $r_\mathrm{s}$ is the scale radius of the NFW density profile;
\item[O] a model with anisotropy of opposite sign at the center and at large radii,
\be 
\beta_{\text{O}}(r)=\theta_{\beta}\dfrac{r-r_\mathrm{s}}{r+r_\mathrm{s}}; 
\ee
\item[M{\L}] the \citet{2005Mamon} profile, 
\be
\beta_{\text{M{\L}}}(r)= 0.5\dfrac{r}{r+\theta_{\beta}};
\ee
\item[OM] the Osipkov-Merritt anisotropy profile \citep{1979Osipkov,1985Merritt},
\be
\beta_{\text{OM}}(r)=\dfrac{r^{2}}{r^{2}+\theta_{\beta}^{2}} .
\ee
\end{enumerate}

Summarizing, we run MAMPOSSt with 3 free parameters: the virial radius $r_{200}$, the scale radius $r_{\text{s}}$ of the mass distribution, and a velocity anisotropy parameter $\theta_{\beta}$. This parameter represents the usual $\beta=1-(\sigma_{\theta}^{2}/\sigma_{r}^{2})$ for the first three models (C, T, O), while for the $\text{M{\L}}$ and OM models it defines a characteristic radius $\theta_\beta=r_\beta$. 
The maximum likelihood solutions are obtained using the NEWUOA software \citep{newuoa} and are shown in Section~\ref{sec:results}.


\section{Cluster Data}
\label{sec:data}

In this section we present the cluster sample, spectroscopic data, and the method for constructing composite clusters.

\begin{figure}
\vskip-0.10in
\centering 
\includegraphics[scale=0.60]{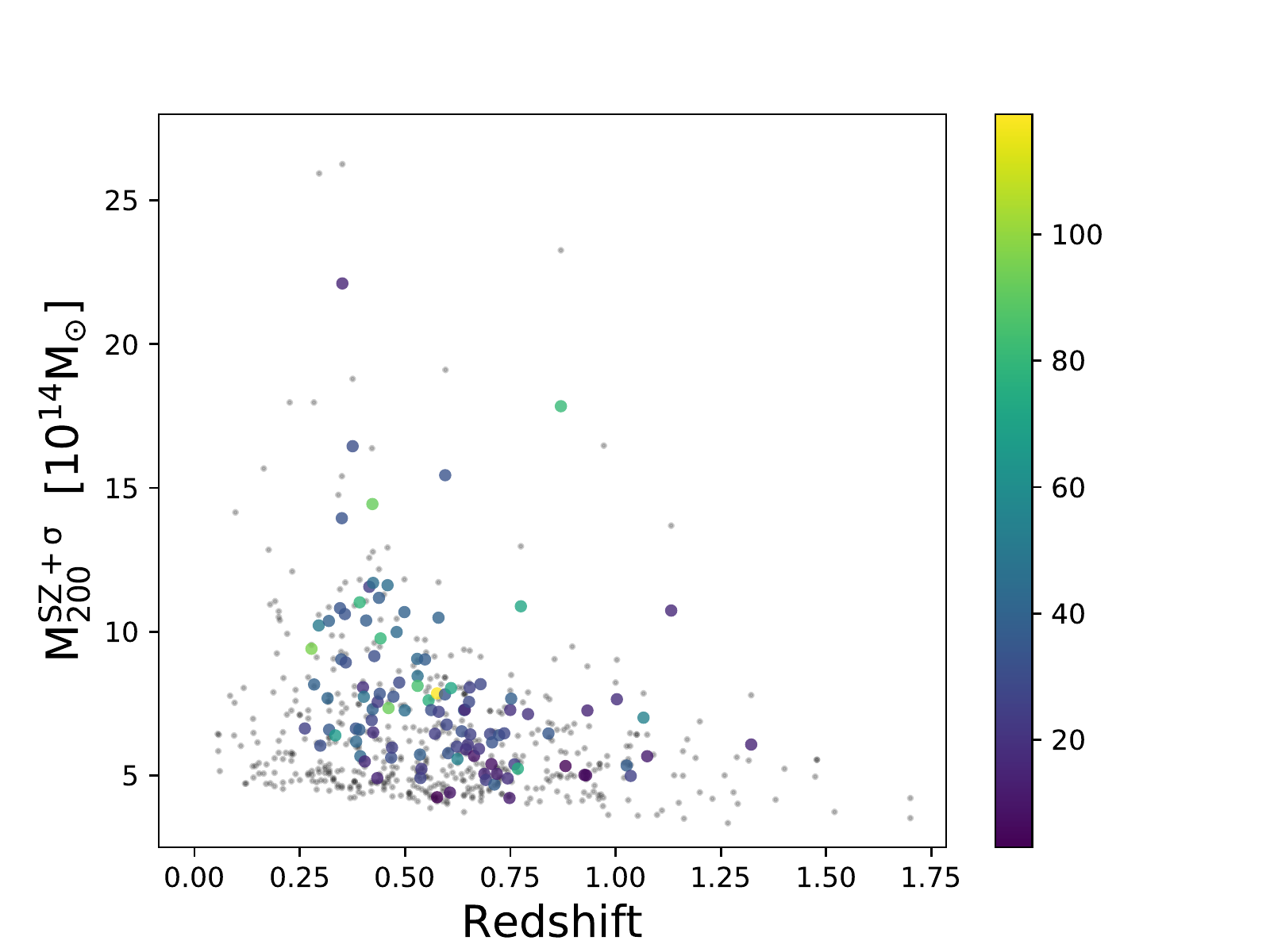}
\vskip-0.13in
\caption {Masses vs redshifts of the cluster sample. The colored dots are coded according to the number of member galaxies in each cluster (see color bar on right). Overplotted in grey is the full distribution of the $2500 \deg^{2}$ SPT-SZ sample. }
\label{fig:zMplot}
\vskip-0.20in
\end{figure}

\subsection{Cluster sample}
\label{sec:spt}

The cluster sample analyzed in this study consists of galaxy clusters detected with the South Pole Telescope (SPT), a 10-meter telescope located within 1 km of the geographical South Pole, observing in three mm-wave bands centered at 95, 150 and 220 GHz \citep[see][]{2011Carlstrom}.  The SPT-SZ survey, imaging 2500 $\text{deg}^{2}$ of the southern sky, has produced data that are used to select galaxy clusters via their thermal SZE signature in the 95 and 150 GHz maps.  This SZE signature arises through the inverse Compton scattering of the Cosmic Microwave Background (CMB) photons and the hot intracluster medium \citep[ICM;][]{sunyaev72}.  We refer the reader to \citet{schaffer11} for details on the survey strategy and data processing.  All of the clusters studied in this work have a high detection significance \citep[$\xi > 4.8$;][]{song12b,bleem15} and have spectroscopic redshifts.

A major advantage of selecting clusters with the SZE rather than other cluster observables lies in the fact that the surface brightness of the SZE signature is independent of the cluster redshift, which together with the expected temperature and density evolution of the cluster ICM at fixed virial mass and the changing solid angle of clusters with redshift  lead an SZE signal to noise selected sample to be approximately mass-limited \citep{haiman01,2001Holder}. 

Because we plan to carry out a dynamical analysis, the cluster sample we analyze includes only those SPT systems with spectroscopic follow-up.  This spectroscopic subsample is not a signal to noise selected sample, but rather is the largest sample of SPT selected clusters we could assemble for the analysis.  We plot the sample of 110 clusters in the space of redshift versus mass in Fig.~\ref{fig:zMplot}, together with the full distribution of the $2500 \deg^{2}$ SPT-SZ sample with the latest available redshifts \citep[][Bocquet et al., in prep]{2016Bayliss,2018Khullar,2018Strazzullo}. Note that in comparison to the SPT-SZ cosmology sample \citep{2016deHaan}, the median redshift and mass for our sample is 0.56 and $7.26 \times 10^{14} \mathrm{M_{\odot}}$ as compared to 0.55  and $6.08 \times 10^{14} \mathrm{M_{\odot}}$.  Approximately 6~percent of the sample lies at $z>1$ as compared to $\sim9$~percent of the full SPT-SZ $\xi>5$ sample (with updated redshifts from Bocquet et al., in prep).  Thus, in comparison to the complete SPT-SZ sample, the subset of those clusters we study here have somewhat higher masses and are somewhat underrepresented at high redshift.  Full information on the cluster sample is provided in Table~\ref{tab:Dataset}.  From left to right  the columns correspond to the SPT cluster designation, the total number of passive galaxy redshifts available, the number of passive galaxy member redshifts, the cluster redshift, the SZE based mass $M_{200}^\mathrm{SZ+\sigma}$ and the corresponding virial radius $R_{200}^\mathrm{SZ+\sigma}$ (described in Section~\ref{sec:SPTmasses}).

\begin{figure}
\vskip-0.05in
\centering 
\includegraphics[scale=0.60]{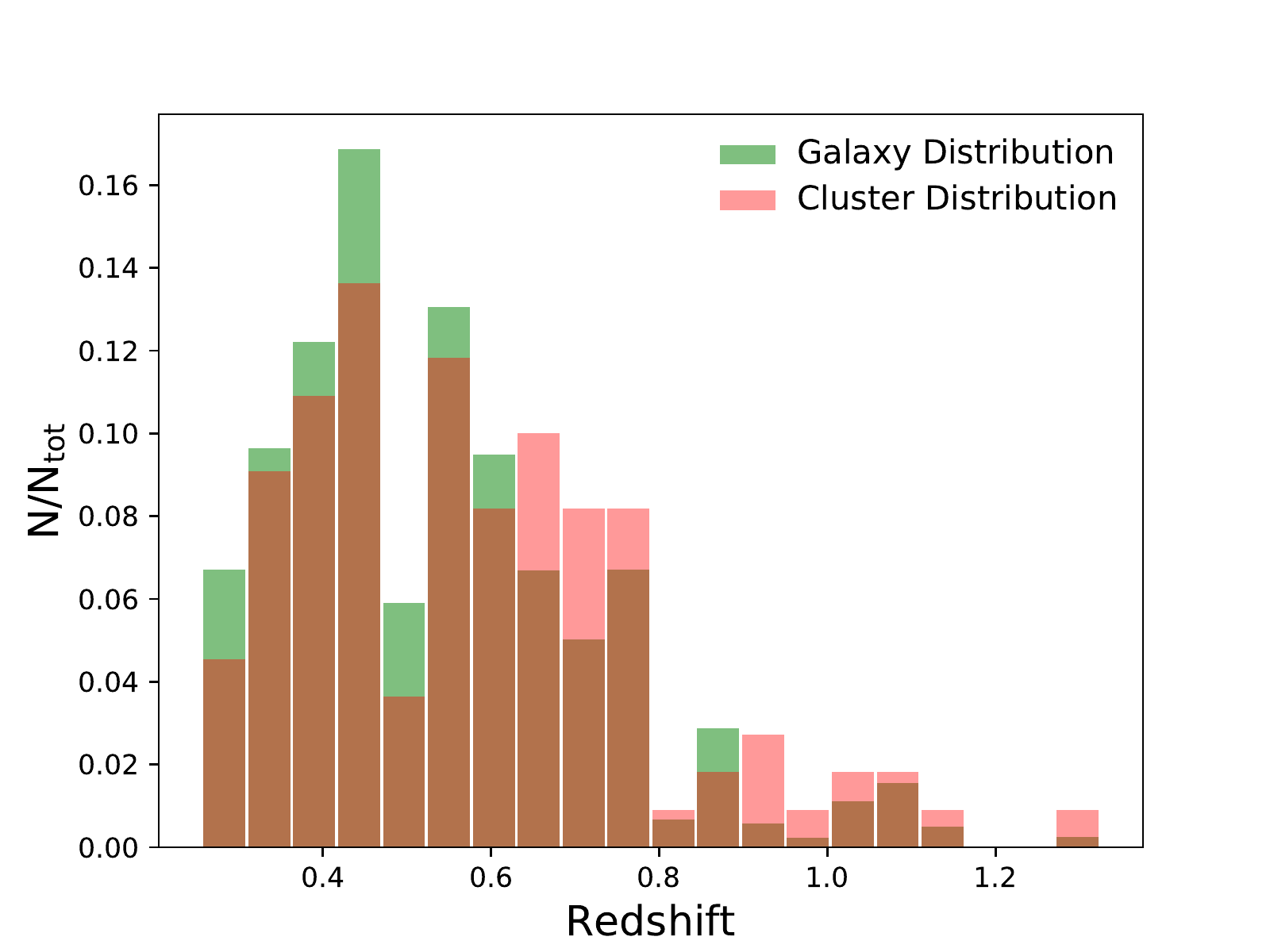}
\vskip-0.15in
\caption {Redshift distribution of member galaxies (green) and clusters (red) in our sample, normalized to the total number of objects in each case.  The two distributions are similar, but a clear trend to have fewer galaxies per cluster at high redshift is visible.}
\label{fig:histo}
\vskip-0.2in
\end{figure}

\subsection{Spectroscopic sample}
\label{sec:gemini}

We use spectroscopic follow-up including data taken using the Gemini Multi Object Spectrograph \citep[GMOS;][]{Hook2004} on Gemini South, the Focal Reducer and low dispersion Spectrograph \citep[FORS2;][]{Appenzeller1998} on VLT Antu, the Inamori Magellan Areal Camera and Spectrograph \citep[IMACS;][]{Dressler2006} on Magellan Baade, and the Low Dispersion Survey Spectrograph \citep[LDSS3;][] {Allington-Smith1994} on Magellan Clay. 

We combine the datasets presented in \citet{2014Ruel} and in \citet{2016Bayliss} with the one described in \citet{Sifon2013}, obtaining a sample of 4695 redshifts. 
In what follows, we split the spectroscopic galaxy sample according to the spectral features into two subsamples of those with emission lines (EL) and those without emission lines (nEL), adopting the determinations presented in  the aforementioned analyses.

Due to the observational strategy of these surveys, which targeted red-sequence (RS) galaxies, the number of nEL galaxies (3834) greatly exceeds the numer of EL galaxies.  Because EL and nEL galaxies are characterized by different spatial and kinematic properties \citep[see, e.g.,][]{mohr96,1997A&A...321...84B, 1999Dressler, 2002Biviano, 2016Biviano, 2017Bayliss,2017Hennig}, in the analysis that follows we focus only on the passive nEL population, for which we have sufficient statistics to properly carry out the dynamical analysis.
In Fig.~\ref{fig:histo} we show the normalized distribution of galaxies and clusters as a function of redshift.

\begin{table}
\caption[]{Cluster spectroscopic sample: We present the cluster name, total number of passive galaxies in the spectroscopic sample $N_\mathrm{tot}$, members used in our analysis after interloper rejection $N_\mathrm{mem}$, cluster redshift $z_\mathrm{clus}$, and cluster mass \Mszsigma\  and  corresponding virial radius $R_{200}^\mathrm{SZ+\sigma}$ from \citet{bocquet15}.}
\small
\begin{tabular}{lcccrc}
\hline\\[-7pt]
Cluster & $N_{\text{tot}}$ & $N_{\text{mem}}$ & $z_{\text{clus}}$  & \multicolumn{1}{c}{\Mszsigma} & $R_{200}^\mathrm{SZ+\sigma} $ \\[2pt]
&  &  &  & \multicolumn{1}{c}{$ [10^{14} \text{M}_{\odot}]$} & [Mpc] \\[2pt]
\hline\\[-7pt]
SPT-CL~J0000-5748   &              27   &              24   &    0.702   &    6.44   &    1.37   \\  
SPT-CL~J0013-4906   &              37   &              36   &    0.408   &    10.39   &    1.81   \\  
SPT-CL~J0014-4952   &              38   &              27   &    0.752   &    7.67   &    1.42   \\  
SPT-CL~J0033-6326   &              28   &              14   &    0.599   &    6.77   &    1.45   \\  
SPT-CL~J0037-5047   &              37   &              17   &    1.026   &    5.34   &    1.13   \\  
SPT-CL~J0040-4407   &              33   &              33   &    0.350   &    13.95   &    2.04   \\  
SPT-CL~J0102-4603   &              35   &              16   &    0.841   &    6.46   &    1.30   \\  
SPT-CL~J0102-4915   &              81   &              80   &    0.870   &    17.84   &    1.80   \\  
SPT-CL~J0106-5943   &              35   &              26   &    0.348   &    9.04   &    1.76   \\  
SPT-CL~J0118-5156   &               8   &               7   &    0.705   &    5.40   &    1.29   \\  
SPT-CL~J0123-4821   &              26   &              18   &    0.655   &    6.43   &    1.40   \\  
SPT-CL~J0142-5032   &              28   &              23   &    0.679   &    8.17   &    1.50   \\  
SPT-CL~J0200-4852   &              45   &              34   &    0.499   &    7.26   &    1.55   \\  
SPT-CL~J0205-5829   &              15   &               8   &    1.322   &    6.07   &    1.06   \\  
SPT-CL~J0205-6432   &              19   &              12   &    0.744   &    4.90   &    1.23   \\  
SPT-CL~J0212-4657   &              26   &              20   &    0.654   &    8.06   &    1.51   \\  
SPT-CL~J0232-5257   &              77   &              61   &    0.556   &    7.61   &    1.54   \\  
SPT-CL~J0233-5819   &               9   &               9   &    0.664   &    5.68   &    1.33   \\  
SPT-CL~J0234-5831   &              24   &              21   &    0.415   &    11.56   &    1.87   \\  
SPT-CL~J0235-5121   &              96   &              82   &    0.278   &    9.41   &    1.84   \\  
SPT-CL~J0236-4938   &              66   &              63   &    0.334   &    6.39   &    1.58   \\  
SPT-CL~J0240-5946   &              19   &              17   &    0.400   &    8.06   &    1.67   \\  
SPT-CL~J0243-4833   &              39   &              37   &    0.498   &    10.68   &    1.76   \\  
SPT-CL~J0243-5930   &              32   &              25   &    0.634   &    6.53   &    1.42   \\  
SPT-CL~J0252-4824   &              27   &              22   &    0.421   &    6.92   &    1.57   \\  
SPT-CL~J0254-5857   &              37   &              32   &    0.438   &    11.18   &    1.83   \\  
SPT-CL~J0257-5732   &              15   &              14   &    0.434   &    4.91   &    1.39   \\  
SPT-CL~J0304-4401   &              45   &              35   &    0.458   &    11.62   &    1.84   \\  
SPT-CL~J0304-4921   &              79   &              72   &    0.392   &    11.02   &    1.85   \\  
SPT-CL~J0307-6225   &              26   &              17   &    0.580   &    7.21   &    1.49   \\  
SPT-CL~J0310-4647   &              33   &              28   &    0.707   &    6.15   &    1.35   \\  
SPT-CL~J0317-5935   &              25   &              18   &    0.469   &    5.97   &    1.47   \\  
SPT-CL~J0324-6236   &              19   &               9   &    0.750   &    7.28   &    1.40   \\  
SPT-CL~J0330-5228   &              80   &              71   &    0.442   &    9.77   &    1.75   \\  
SPT-CL~J0334-4659   &              29   &              25   &    0.486   &    8.23   &    1.62   \\  
SPT-CL~J0346-5439   &              85   &              79   &    0.530   &    8.11   &    1.59   \\  
SPT-CL~J0348-4515   &              31   &              24   &    0.359   &    8.93   &    1.75   \\  
SPT-CL~J0352-5647   &              22   &              16   &    0.649   &    6.07   &    1.37   \\  
SPT-CL~J0356-5337   &              26   &               5   &    1.036   &    4.98   &    1.10   \\  
SPT-CL~J0403-5719   &              31   &              24   &    0.467   &    5.62   &    1.44   \\  
SPT-CL~J0406-4805   &              28   &              26   &    0.736   &    6.46   &    1.35   \\  
SPT-CL~J0411-4819   &              45   &              42   &    0.424   &    11.70   &    1.87   \\  
SPT-CL~J0417-4748   &              40   &              30   &    0.579   &    10.49   &    1.69   \\  
SPT-CL~J0426-5455   &              15   &              11   &    0.642   &    7.28   &    1.46   \\  
SPT-CL~J0433-5630   &              24   &              18   &    0.692   &    4.84   &    1.25   \\  
SPT-CL~J0438-5419   &              92   &              87   &    0.422   &    14.44   &    2.00   \\  
SPT-CL~J0449-4901   &              20   &              16   &    0.792   &    7.13   &    1.37   \\  
SPT-CL~J0456-5116   &              31   &              20   &    0.562   &    7.27   &    1.51   \\  
SPT-CL~J0509-5342   &              93   &              88   &    0.461   &    7.34   &    1.58   \\  
SPT-CL~J0511-5154   &              18   &              14   &    0.645   &    5.90   &    1.36   \\  
SPT-CL~J0516-5430   &              51   &              47   &    0.295   &    10.22   &    1.88   \\  
SPT-CL~J0521-5104   &              21   &              21   &    0.675   &    5.92   &    1.35   \\  
SPT-CL~J0528-5300   &              75   &              63   &    0.768   &    5.24   &    1.25   \\  
SPT-CL~J0533-5005   &               4   &               4   &    0.881   &    5.33   &    1.20   \\  
SPT-CL~J0534-5937   &               3   &               3   &    0.576   &    4.24   &    1.25   \\  
SPT-CL~J0540-5744   &              24   &              17   &    0.760   &    5.38   &    1.26   \\  
SPT-CL~J0542-4100   &              36   &              29   &    0.640   &    7.27   &    1.46   \\ 
SPT-CL~J0546-5345   &              54   &              49   &    1.066   &    7.01   &    1.22   \\
SPT-CL~J0549-6205   &              31   &              26   &    0.375   &    16.45   &    2.13   \\  
SPT-CL~J0551-5709   &              39   &              30   &    0.423   &    7.30   &    1.60   \\  [2pt]
\hline
\end{tabular}
\label{tab:Dataset}
\end{table}

\begin{table}
\caption*{{\bf Table~\ref{tab:Dataset}:} Cluster spectroscopic sample - Continued.}
\small
\begin{tabular}{lcccrc}
\hline\\[-7pt]
Cluster & $N_{\text{tot}}$ & $N_{\text{mem}}$ & $z_{\text{clus}}$  & \multicolumn{1}{c}{\Mszsigma} & $R_{200}^\mathrm{SZ+\sigma}$  \\[2pt]
&  &  &  & \multicolumn{1}{c}{$ [10^{14} \text{M}_{\odot}]$} & [Mpc] \\ [2pt]
\hline \\[-7pt]
SPT-CL~J0555-6406   &              34   &              27   &    0.345   &    10.82   &    1.88   \\ 
SPT-CL~J0559-5249   &              72   &              67   &    0.609   &    8.04   &    1.53   \\  
SPT-CL~J0655-5234   &              33   &              30   &    0.472   &    7.74   &    1.60   \\  
SPT-CL~J2017-6258   &              38   &              35   &    0.535   &    5.73   &    1.41   \\  
SPT-CL~J2020-6314   &              27   &              18   &    0.537   &    4.91   &    1.34   \\  
SPT-CL~J2022-6323   &              34   &              29   &    0.383   &    6.63   &    1.57   \\  
SPT-CL~J2026-4513   &              15   &              11   &    0.688   &    5.05   &    1.27   \\  
SPT-CL~J2030-5638   &              42   &              36   &    0.394   &    5.67   &    1.48   \\  
SPT-CL~J2032-5627   &              39   &              32   &    0.284   &    8.16   &    1.75   \\  
SPT-CL~J2035-5251   &              42   &              29   &    0.529   &    9.05   &    1.65   \\  
SPT-CL~J2040-5725   &               7   &               4   &    0.930   &    5.00   &    1.15   \\  
SPT-CL~J2043-5035   &              33   &              21   &    0.723   &    6.40   &    1.36   \\  
SPT-CL~J2056-5459   &              13   &              11   &    0.718   &    5.06   &    1.26   \\  
SPT-CL~J2058-5608   &              10   &               6   &    0.607   &    4.40   &    1.25   \\  
SPT-CL~J2100-4548   &              37   &              18   &    0.712   &    4.68   &    1.23   \\  
SPT-CL~J2106-5844   &              16   &              16   &    1.132   &    10.74   &    1.37   \\  
SPT-CL~J2115-4659   &              31   &              28   &    0.299   &    6.03   &    1.57   \\  
SPT-CL~J2118-5055   &              55   &              33   &    0.624   &    5.57   &    1.35   \\  
SPT-CL~J2124-6124   &              23   &              21   &    0.435   &    7.56   &    1.61   \\  
SPT-CL~J2130-6458   &              41   &              40   &    0.316   &    7.69   &    1.69   \\  
SPT-CL~J2135-5726   &              31   &              30   &    0.427   &    9.15   &    1.72   \\  
SPT-CL~J2136-4704   &              19   &              19   &    0.424   &    6.50   &    1.53   \\  
SPT-CL~J2136-6307   &               6   &               6   &    0.926   &    5.02   &    1.15   \\  
SPT-CL~J2138-6008   &              39   &              32   &    0.319   &    10.37   &    1.87   \\  
SPT-CL~J2140-5727   &              17   &              11   &    0.404   &    5.49   &    1.46   \\  
SPT-CL~J2145-5644   &              43   &              35   &    0.480   &    9.99   &    1.73   \\  
SPT-CL~J2146-4633   &              15   &               7   &    0.933   &    7.26   &    1.30   \\  
SPT-CL~J2146-4846   &              26   &              25   &    0.623   &    6.00   &    1.38   \\  
SPT-CL~J2146-5736   &              34   &              23   &    0.602   &    5.77   &    1.38   \\  
SPT-CL~J2148-6116   &              24   &              24   &    0.571   &    6.45   &    1.45   \\  
SPT-CL~J2155-6048   &              23   &              19   &    0.539   &    5.23   &    1.36   \\  
SPT-CL~J2159-6244   &              38   &              36   &    0.391   &    6.59   &    1.56   \\  
SPT-CL~J2222-4834   &              29   &              25   &    0.652   &    7.56   &    1.48   \\  
SPT-CL~J2232-5959   &              34   &              26   &    0.595   &    7.82   &    1.53   \\  
SPT-CL~J2233-5339   &              33   &              28   &    0.440   &    7.84   &    1.62   \\  
SPT-CL~J2248-4431   &              15   &              14   &    0.351   &    22.11   &    2.37   \\  
SPT-CL~J2300-5331   &              25   &              21   &    0.262   &    6.63   &    1.64   \\  
SPT-CL~J2301-5546   &              12   &               8   &    0.748   &    4.21   &    1.17   \\  
SPT-CL~J2306-6505   &              46   &              42   &    0.530   &    8.46   &    1.61   \\  
SPT-CL~J2325-4111   &              33   &              27   &    0.357   &    10.61   &    1.86   \\  
SPT-CL~J2331-5051   &             119   &             108   &    0.576   &    7.85   &    1.54   \\  
SPT-CL~J2332-5358   &              47   &              45   &    0.402   &    7.74   &    1.64   \\  
SPT-CL~J2335-4544   &              37   &              33   &    0.547   &    9.04   &    1.63   \\  
SPT-CL~J2337-5942   &              28   &              19   &    0.775   &    10.88   &    1.59   \\  
SPT-CL~J2341-5119   &              18   &              13   &    1.003   &    7.65   &    1.29   \\  
SPT-CL~J2342-5411   &              12   &               7   &    1.075   &    5.67   &    1.14   \\  
SPT-CL~J2344-4243   &              33   &              25   &    0.595   &    15.44   &    1.91   \\  
SPT-CL~J2351-5452   &              42   &              30   &    0.384   &    6.18   &    1.53   \\  
SPT-CL~J2355-5055   &              36   &              33   &    0.320   &    6.59   &    1.61   \\  
SPT-CL~J2359-5009   &              44   &              37   &    0.775   &    5.17   &    1.24   \\[2pt]  
\hline
\end{tabular}
\end{table}

\subsection{Construction of composite clusters}
\label{sec:stack}

To enable a precise determination of the cluster masses and the velocity anisotropy profiles, we cannot rely on the few spectroscopic members in the individual clusters of our sample. For this reason, we either create composite clusters with much more dynamical information or, as for the results presented in Section~\ref{sec:masstension}, fit to a common model across the cluster ensemble by combining the likelihoods associated with each individual cluster. 
The composite cluster approach has been adopted in previous analyses \citep{1997Carlberg,vanderMarel2000, Katgert2004, 2009Biviano}, and it is supported by cosmological simulations that predict cosmological halos can be characterized by a universal mass density profile with a concentration that depends mildly  on the halo mass \citep{NA97.1}.  We return to the method using a combination of individual cluster likelihoods in Section~\ref{sec:masstension}.

We create composite clusters by combining cluster subsamples within different mass and redshift ranges in such a way that each subsample includes an adequate number of members.  Considering that tests done using MAMPOSSt on cosmological simulations indicate that with 500 tracers the code is able to recover the mass and anisotropy parameters with a suitably small uncertainty \citep{2013MAMPOSSt}, we create similarly sized subsamples.
With this approach we can construct 5 composite clusters selected either as redshift or mass subsamples.

Table~\ref{tab:bins} lists the characteristics of the composite clusters created within redshift bins.  
From left to right the columns correspond to the bin number, the redshift range of the clusters combined to create the composite system in that bin, the mean redshift of these clusters, the number of clusters, the total number of galaxy redshifts, the number of cluster members with redshifts, and the average SPT based mass of the individual clusters in the sample.  In the following subsections we describe in detail how these composite clusters are created and how interlopers are removed from the spectroscopic samples.

\begin{table}
\centering
\caption{Characteristics of the composite clusters in redshift: columns show the bin number, redshift range, mean redshift, number of included clusters $N_{\text{clus}} $, number of total galaxies $N_{\text{tot}}$, number of member galaxies $N_{\text{mem}}$ and mean SPT based mass $\langle$\Mszsigma$\rangle$.}
\begin{tabular}{cccccccc}
\hline\\[-7pt]
Bin & Redshift & $\langle z \rangle$ & $N_{\text{clus}} $ & $N_{\text{tot}}$ & $N_{\text{mem}}$ & $\langle$\Mszsigma$\rangle$ \\[2pt]
& range&&&&  & $[10^{14} M_{\odot}]$  \\[2pt] 
\hline\\[-7pt]
1 & 0.26-0.38 &  0.33 &  18 &  712  & 593 & 9.23 \\
2 & 0.39-0.44 & 0.42 & 19 & 744 &  644 & 9.50 \\
3 & 0.46-0.56 & 0.51 & 17 & 758 &  615 & 7.98 \\
4 & 0.56-0.71 & 0.62 & 30 & 904 & 675 & 7.31 \\
5 & 0.71-1.32 & 0.86 & 26 & 716 &  459 & 7.60 \\
- & 0.26-1.32 & 0.55 & 110 & 3834 &  2966 & 8.28 \\
\hline
\end{tabular}
\label{tab:bins}
\end{table}

\subsubsection{Rescaling observables with SZE based mass estimates}
\label{sec:SPTmasses}

To create composite clusters we choose the Brightest Cluster Galaxy (BCG) as the cluster center.  We determine these positions using results from previous and ongoing work \citep[][Stalder et al., in prep]{song12b,Sifon2013}, allowing us to calculate a projected separation $R$ for each galaxy from the cluster center. 
We extract the rest frame LOS velocity $\mathrm{v_\text{rf}}$ from the galaxy redshift $z$ and equivalent velocity $\mathrm{v}(z)$ as $\mathrm{v}_{\text{rf}} \equiv (\mathrm{v}(z) - \mathrm{v}(z_\text{c}))/(1+z_{\text{c}})$, where $z_\text{c}$ is the cluster redshift.
 
Because our clusters span a wide range of mass and redshift, we must scale the galaxy projected cluster distances $R$ and LOS velocities to the values of a fiducial cluster mass using an estimate of the individual cluster virial radius and velocity. For convenience, in the composite cluster we use the mean mass $\left<M_{200}\right>$ and the mean redshift $\left< z\right>$ of the ensemble of clusters in that bin to calculate the associated $\left<R_{200}\right>$ and $\left<\mathrm{v}_{200}\right>$ that we adopt as fiducial values for the bin.  
Thus, the rescaled observables for each galaxy $i$ within a cluster $j$ in a particular bin are $R_{\text{i,j}}= R_\mathrm{i}\, \left<R_{200}\right>/R_\mathrm{200,j}$ and $\mathrm{v_{\text{i,j}}= v_{\text{rf,i}}\, \left<v_{200}\right>/v_\mathrm{200,j}}$, where $R_\mathrm{200,j}$ and $\mathrm{v_\mathrm{200,j}}$ are the virial radius and circular velocity, respectively, of cluster $j$.  The circular velocity is defined using the virial condition $\mathrm{v}_{200}^2=G M_{200}/R_{200}$.  

To perform the normalization, we need precise estimates of the cluster mass.  Given that the SPT sample is SZE selected, we adopt the SZE observable $\xi$, which is the detection signal to noise and is correlated to the underlying cluster virial mass \citep{andersson11,benson13}, as the source of our cluster mass estimates.  SPT masses derived from $\xi$ have a statistical uncertainty that depends on the intrinsic scatter in the $\xi$--mass relation and on the observational uncertainty in the signal to noise of the detection; together, these lead characteristically to $\sim$20~percent statistical uncertainty in the cluster virial mass.  In addition, there are systematic uncertainties remaining from the calibration of the mass--observable relation that are currently at the $\sim$15~percent level.  As discussed further below, the systematic mass scale uncertainties at this level have no impact on our analysis, but the cluster to cluster statistical mass uncertainty of $\sim$20~percent introduces a corresponding uncertainty in $R_{200}$ and $v_{200}$ of $\sim$7~percent.

The mass--observable relation for the SPT sample can be calibrated in different ways, and in the current analysis we use two different approaches. The first approach uses direct cluster mass measurements from weak lensing, velocity dispersions, X-ray measurements to calibrate the SZE-mass observable relation.  The second approach fits the SPT cluster distribution in signal to noise $\xi$ and redshift $z$ and the mass-observable relation to a flat $\Lambda$CDM model with external cosmological priors from other datasets.  These constraints from, e.g., CMB anisotropy, baryon acoustic oscillations, etc., effectively constrain the mass-observable relation so that the cluster mass function implied by SPT cluster distribution in $\xi$ and $z$ is consistent with that expected given the external cosmological priors. 
As an example \citep[see][]{bocquet15}, the inclusion of external cosmological constraints from Planck CMB in the SPT mass--observable calibration leads to a shift of cluster masses to $\sim$25~percent higher values in comparison to the case where the mass calibration is undertaken with direct cluster mass measurements from velocity dispersions.

Given the need to have an initial mass estimate to enable the stacking of the clusters for a dynamical analysis, we carry out the analysis with initial SPT based masses calibrated in two different ways.   The first uses the SPT cluster counts together with direct cluster mass information from velocity dispersions.  Here we refer to these masses as \Mszsigma.  The second uses the SPT cluster counts together with external cosmological priors from Planck CMB anisotropy and distance measurements (BAO, Supernovae), specifically we use mass-calibration inferred from the SPTCL +Planck +WP+BAO+SNIa data set in \citet{bocquet15}. 
We refer to these masses as \MszPlanck.  In both cases the SPT mass--observable relation is calibrated using $M_{500}$ \citep{bocquet15}, and so we transform from $M_{500}$ by assuming an NFW model with a concentration $c_{200}$ sampled from cosmological $N$-body simulations \citep{duffy08}.  The $r_{200}$ and $\mathrm{v}_{200}$ are then easily obtained, depending only on the cluster redshift and the adopted cosmology.  

As discussed in Section~\ref{sec:compositemasses}, the dynamical mass measurements of the composite clusters have only a weak dependence on the mass scale of the velocities and radii used to build the composite clusters;  specifically, the $\sim$25~percent shift between the masses \Mszsigma\ and \MszPlanck\ used to build the two sets of composite clusters has no significant impact on the final dynamical masses.

Because we are carrying out a Jeans analysis, which is based on the assumption of dynamical equilibrium, we restrict our analysis to the cluster virial region ($R \leq r_{200}$). Moreover, we exclude the very central cluster region ($R \leq$ 50 kpc).  In fact, the composite clusters are centered on the BCG, which we exclude from the dynamical analysis.  We note also that in the composite clusters the characteristic asphericity of individual clusters is averaged down, leading to a combined system that is approximately spherical in agreement with the dynamical model we are employing.

\begin{figure}
\vskip-0.1in
\includegraphics[scale=0.55]{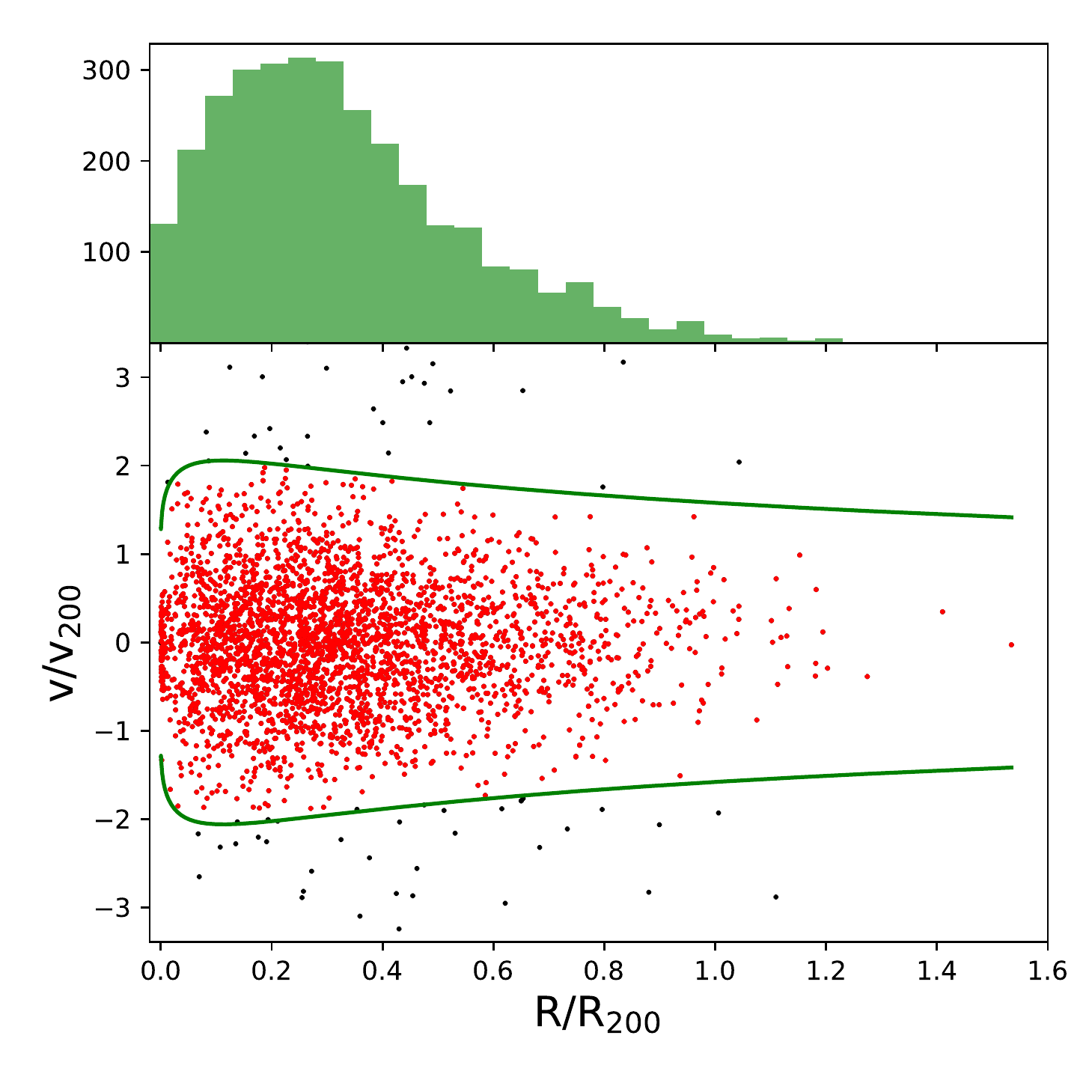}
\vskip-0.15in
\caption {The projected phase-space diagram (below) in the rescaled coordinates $(R/R_{200},\mathrm{v/v_{200}})$ for the full sample.  Green lines represent the radially-dependent 2.7$\sigma_{\text{LOS}}$ cut used to reject interlopers (indicated by black dots). The radial distribution of the member galaxies with measured redshifts is shown in the upper panel.}
\label{fig:interlop_general}
\vskip-0.2in
\end{figure}

\subsubsection{Interloper rejection}
\label{sec:interl_rej}

One benefit of constructing composite clusters is that we can more easily identify and reject some interloper galaxies, i.e., galaxies that are projected inside the cluster virial region, but do not actually lie inside it. We do so by using the ``Clean" method \citep{2013MAMPOSSt}, which is based on the identification of the cluster members on the basis of their projected phase-space location ($R, \mathrm{v}$).  The LOS velocity dispersion $\sigma_{\text{LOS}}$ of the composite cluster is used to estimate the cluster mass using a scaling relation calibrated using numerical simulations
\citep[e.g.,][]{saro13}, and an NFW mass profile with concentration sampled from the theoretical mass-concentration relation of \citet{Maccio2008}. Thereafter, assuming the M\L\ velocity anisotropy profile model, and given the $M(r)$ of the cluster, an LOS velocity dispersion profile $\sigma_{\text{LOS}}(R)$ is calculated and used to iteratively reject galaxies with $| \mathrm{v}_\text{rf} |  > 2.7 \sigma_{\text{LOS}}$ at any clustercentric distance \citep[see][]{Mamon2010, 2013MAMPOSSt}. 
As an example, we present in Fig.~\ref{fig:interlop_general} the location of galaxies in projected phase-space with the identification of cluster member galaxies for the full sample.

We note that even after cleaning the dynamical sample, it is still contaminated to some degree by interlopers.  One reason for this is that galaxies near the cluster turn-around radius (i.e., well outside the virial radius) will have small LOS velocities.  Therefore, if those galaxies are projected onto the cluster virial region they simply cannot be separated from the galaxies that actually lie within the cluster virial radius.   Analysis of cosmological $N$-body simulations shows that when passive galaxies are selected this contamination is characteristically $\sim20$~percent \citep{saro13} for SPT mass scale clusters and represents a systematic, whose effects on the dynamical analysis require further exploration \citep[see also discussion in][]{2015Old}.

\subsubsection{Galaxy number density profiles}
\label{sec:nR}

We cannot simply adopt the spectroscopic sample to measure the number density profiles of the galaxy population, because the structure of the spectrographs and our observing strategy of visiting each cluster typically with only two masks will generally result in a radially dependent incompleteness.
This incompleteness, if not accounted for, would affect the determination of the cluster projected number density profile $\Sigma(R)$, related to the 3D number density profile $\nu(r)$ through the Abel inversion \citep{1987Binney}. As only the logarithmic derivative of $\nu(r)$ enters the Jeans equation (see equation~(\ref{eqn:jeans2})), the absolute normalization of the galaxy number density profile has no impact on our analysis. However, a radially dependent incompleteness in the velocity sample would impact our analysis.

Thus, for our analysis we rely on a study of the galaxy populations in 74 SZE selected clusters from the SPT survey that were imaged as part of the Dark Energy Survey (DES) Science Verification phase (H17). This complete sky-area-selected subsample of the SPT-SZ cluster sample has a redshift and mass distribution that are consistent with our sample.  The number density profile of the red sequence population of galaxies is found to be well fit by an NFW model out to radii of $4r_{200}$, with a concentration of cluster galaxies $c_{\text{gal}}=5.37^{+0.27}_{-0.24}$.  The non-red sequence population is much less centrally concentrated, but in our analysis we are focused only on the nEL galaxy populations.  No statistically significant redshift or mass trends in the galaxy concentration were identified.  Therefore, we adopt the number density profile described by an NFW profile with the above-mentioned value of $c_{\text{gal}}$ and a scale radius  $r_{\text{s, gal}}=R_{200}/c_{\text{gal}}$. Implicit in this approach is the assumption that the dynamical properties of our spectroscopic sample are consistent with the dynamical properties of the red sequence galaxy population used to measure the radial profiles.

We test this assumption of a fixed $c_{\text{gal}}$ by including in the Likelihood an additional parameter $r_{\text{s, gal}}$, given by the ratio $R_{200}/c_{\text{gal}}$, where $c_{\text{gal}}$ is given by H17
\begin{equation}
c_{\text{gal}} = A \left( \dfrac{M_{200}}{M_{piv}} \right)^{B} \left( \dfrac{1+z}{1+z_{piv}} \right)^{C} ,
\end{equation}
where, for red sequence galaxies, $A=5.47 \pm 0.53, B=?0.01 \pm 0.10, C=0.15 \pm 0.30$. We choose mass and redshift pivot points $M_{piv} = 5 \times 10^{14}$ and $z_{piv} = 0.53$, respectively, corresponding to the median mass and redshift of our sample. 
With this additional parameter, fitting for A, B, and C with Gaussian priors corresponding to the uncertainties listed above,  we adopt the Markov Chain Monte Carlo (MCMC) method for sampling from the probability distribution, utilizing the \textit{emcee} code \citep{2013emcee}. Our analysis provides values of the scale radius and virial radius consistent within $1\sigma$ of the measured values reported in H17.


\begin{table}
\centering
\begin{threeparttable}
\renewcommand{\TPTminimum}{\linewidth}
\caption{Likelihood ratio (or Bayes factor; see equation~\ref{eq:bayesfactor}) for the NFW, Burkert, Einasto, Hernquist and SIS mass profiles is presented for each anisotropy profile analyzed using the composite cluster constructed using the full dynamical sample.  As noted in the text,
we cannot discard any of the mass models aside from the SIS.}
\small
\begin{tabular}{c|ccccc}
\hline
$\beta(r)$ model  & \multicolumn{5}{c}{Mass model}\\ [3pt]
\hline\\[-7pt]
  & NFW & Burkert & Einasto  & Hernquist & SIS \\ 
\hline\\[-7pt]
\multicolumn{1}{c|} {C } & 0.52 & 0.19 & 0.47  &  1.00 & $9.04 \times 10^{-4}$ \\ [3pt] 
\multicolumn{1}{c|} {M\L}& 0.68  &0.07  & 1.00  &  0.89 & $7.46 \times 10^{-4}$ \\ [3pt] 
\multicolumn{1}{c|} {OM } & 0.45 & 0.18 & 0.46   & 1.00 & $4.44 \times 10^{-6}$  \\ [3pt] 
\multicolumn{1}{c|} {O }& 0.90 & 0.03 & 1.00  &  0.43  & \\ [3pt] 
\multicolumn{1}{c|} {T } & 0.75 & 0.06 & 0.84   &  1.00 &  \\ [3pt] 
\hline
\end{tabular}
\label{tab:massres}
\end{threeparttable}
\vskip-0.0in
\end{table}

\begin{table}
\centering
\begin{threeparttable}
\renewcommand{\TPTminimum}{\linewidth}
\caption{Parameter constraints from the MAMPOSSt analysis of the composite clusters defined in Table~\ref{tab:bins}. Columns represent the velocity anisotropy model $\beta(r)$, the virial radius $r_{200}$, the scale radius $r_\mathrm{s}$, the anisotropy parameter $\theta_{\beta}$ and the Bayes factor from equation~\ref{eq:bayesfactor}. For each composite cluster, we also report the value of $r_{\nu}$, which is the scale radius of the galaxy number density profile, obtained from the ratio between the mean value of the $R_{200}^\mathrm{SZ+\sigma}$ in each redshift bin, and the fixed $c_{gal}$ value (see Section~\ref{sec:nR}). Note that for the anisotropy models C, T, and O the anisotropy parameter does not have units, while for the \text{M{\L}} and OM the values are evaluated in Mpc. }
\small
\begin{tabular}{c|cccc}
\hline
$\beta(r)$   & $r_{200}$ & $r_\mathrm{s}$ & $\theta_{\beta}$  & Bayes \\ 
model  & [Mpc] & [Mpc] &  &  Factor \\[3pt] \hline\\[-7pt]

& \multicolumn{4}{c}{$\langle z \rangle = 0.33 $ ; $r_{\nu}=0.33$ Mpc}\\ [3pt]
\hline\\[-7pt]
\multicolumn{1}{c|} {C } & $ 1.84 _{- 0.07 }^{+ 0.12 }$ & $ 0.34 _{- 0.12 }^{+ 0.54 }$ & $ 0.28 _{- 0.24 }^{+ 0.91 }$  &  0.39 \\ [3pt] 
\multicolumn{1}{c|} {M\L}& $ 1.79 _{- 0.05 }^{+ 0.14 }$ & $ 0.31 _{- 0.08 }^{+ 0.25 }$ & $ 0.37 _{- 0.37 }^{+ 3.99 }$   &  0.94 \\ [3pt] 
\multicolumn{1}{c|} {OM } & $ 1.68 _{- 0.07 }^{+ 0.17 }$ & $ 0.19 _{- 0.05 }^{+ 0.13 }$ & $ 1.97 _{- 0.77 }^{+ 5.19 }$   & 1.00 \\ [3pt] 
\multicolumn{1}{c|} {O }& $ 1.87 _{- 0.14 }^{+ 0.17 }$ & $ 0.24 _{- 0.06 }^{+ 0.15 }$ & $ -0.56 _{- 0.43 }^{+ 2.38 }$  &  0.15 \\ [3pt] 
\multicolumn{1}{c|} {T } & $ 1.81 _{- 0.09 }^{+ 0.09 }$ & $ 0.27 _{- 0.08 }^{+ 0.19 }$ & $ 0.19_{- 0.38}^{+ 2.08 }$   &  0.28 \\ [3pt] 

\hline\\[-7pt]
& \multicolumn{4}{c}{$\langle z \rangle = 0.42 $ ; $r_{\nu}=0.32$ Mpc}\\ [3pt]
\hline\\[-7pt]
\multicolumn{1}{c|} {C } & $ 1.85 _{- 0.07 }^{+ 0.10 }$ & $ 0.44 _{- 0.14 }^{+ 0.53 }$ & $ 0.20 _{- 0.27 }^{+ 0.90 }$   &  0.09 \\ [3pt] 
\multicolumn{1}{c|} {M\L } & $ 1.79 _{- 0.06 }^{+ 0.12 }$ & $ 0.40 _{- 0.11 }^{+ 0.17 }$ & $ 0.86 _{- 0.56 }^{+ 8.79 }$ & 1.00 \\ [3pt] 
\multicolumn{1}{c|} {OM } & $ 1.62 _{- 0.05 }^{+ 0.17 }$ & $ 0.27 _{- 0.06 }^{+ 0.15 }$ & $ 1.19 _{- 0.37 }^{+ 2.97 }$  &  0.25 \\ [3pt] 
\multicolumn{1}{c|} {O } & $ 1.78 _{- 0.10 }^{+ 0.15 }$ & $ 0.35 _{- 0.07 }^{+ 0.12 }$ & $ 0.39_{- 0.41 }^{+ 1.66 }$  & 0.12 \\ [3pt] 
\multicolumn{1}{c|} {T } & $ 1.81 _{- 0.09 }^{+ 0.09 }$ & $ 0.42 _{- 0.10 }^{+ 0.22 }$ & $ 0.39_{- 0.32 }^{+ 1.59 }$ & 0.18 \\ [3pt] 

\hline\\[-7pt]
& \multicolumn{4}{c}{$\langle z \rangle = 0.51 $ ; $r_{\nu}=0.29$ Mpc}\\ [3pt]
\hline\\[-7pt]
\multicolumn{1}{c|} {C }& $ 1.69 _{- 0.05 }^{+ 0.08 }$ & $ 0.32 _{- 0.09 }^{+ 0.23 }$ & $ 0.19_{- 0.26 }^{+ 0.77 }$   & 0.02 \\ [3pt] 
\multicolumn{1}{c|} {M\L } & $ 1.66 _{- 0.05 }^{+ 0.08 }$ & $ 0.33 _{- 0.10 }^{+ 0.13 }$ & $ 0.53 _{- 0.30 }^{+ 5.13 }$   &  0.32 \\ [3pt] 
\multicolumn{1}{c|} {OM} & $ 1.51 _{- 0.04 }^{+ 0.07 }$ & $ 0.33 _{- 0.12 }^{+ 0.19 }$ & $ 0.76 _{- 0.20 }^{+ 1.35 }$ &  0.13 \\ [3pt] 
\multicolumn{1}{c|} {O } & $ 1.56 _{- 0.05 }^{+ 0.06 }$ & $ 0.31 _{- 0.05 }^{+ 0.07 }$ & $ 0.93_{- 0.05 }^{+ 0.21 }$  & 1.00 \\ [3pt] 
\multicolumn{1}{c|} {T } & $ 1.64 _{- 0.06 }^{+ 0.06 }$ & $ 0.40 _{- 0.09 }^{+ 0.17 }$ & $ 0.68_{- 0.18 }^{+ 0.88 }$  & 0.21 \\ [3pt] 

\hline\\[-7pt]
& \multicolumn{4}{c}{$\langle z \rangle = 0.62 $ ; $r_{\nu}=0.27$ Mpc}\\ [3pt]
\hline\\[-7pt]
\multicolumn{1}{c|} {C } & $ 1.53 _{- 0.05 }^{+ 0.06 }$ & $ 0.35 _{- 0.11 }^{+ 0.31 }$ & $ 0.39_{- 0.20 }^{+ 0.64 }$  &  0.27 \\ [3pt] 
\multicolumn{1}{c|} {M\L } & $ 1.52 _{- 0.05 }^{+ 0.05 }$ & $ 0.39 _{- 0.13 }^{+ 0.13 }$ & $ 0.06 _{- 0.05 }^{+ 0.50 }$ & 0.15 \\ [3pt] 
\multicolumn{1}{c|} {OM } & $ 1.43 _{- 0.04 }^{+ 0.09 }$ & $ 0.18 _{- 0.04 }^{+ 0.09 }$ & $ 2.40 _{- 0.81 }^{+ 5.74 }$ & 1.00 \\ [3pt] 
\multicolumn{1}{c|} {O } & $ 1.51 _{- 0.06 }^{+ 0.06 }$ & $ 0.23 _{- 0.05 }^{+ 0.09 }$ & $ -0.02 _{- 0.35 }^{+ 1.77 }$ & 0.09 \\ [3pt] 
\multicolumn{1}{c|} {T } & $ 1.51 _{- 0.06 }^{+ 0.05 }$ & $ 0.27 _{- 0.07 }^{+ 0.15 }$ & $ 0.34_{- 0.30 }^{+ 1.46 }$ &  0.22 \\ [3pt] 

\hline\\[-7pt]
& \multicolumn{4}{c}{$\langle z \rangle = 0.86 $ ; $r_{\nu}=0.25$ Mpc}\\ [3pt]
\hline\\[-7pt]
\multicolumn{1}{c|} {C }& $ 1.30 _{- 0.05 }^{+ 0.06 }$ & $ 0.26 _{- 0.09 }^{+ 0.31 }$ & $ 0.23_{- 0.31 }^{+ 1.04 }$ & 0.12 \\ [3pt] 
\multicolumn{1}{c|} {M\L } & $ 1.28 _{- 0.04 }^{+ 0.07 }$ & $ 0.25 _{- 0.08 }^{+ 0.12 }$ & $ 0.58 _{- 0.38 }^{+ 7.20 }$ & 1.00 \\ [3pt] 
\multicolumn{1}{c|} {OM } & $ 1.19 _{- 0.03 }^{+ 0.09 }$ & $ 0.19 _{- 0.05 }^{+ 0.11 }$ & $ 1.08 _{- 0.44 }^{+ 3.07 }$ &  0.32 \\ [3pt] 
\multicolumn{1}{c|} {O } & $ 1.25 _{- 0.05 }^{+ 0.06 }$ & $ 0.25 _{- 0.05 }^{+ 0.08 }$ & $ 0.68_{- 0.24 }^{+ 1.01 }$ & 0.33 \\ [3pt] 
\multicolumn{1}{c|} {T } & $ 1.28 _{- 0.06 }^{+ 0.05 }$ & $ 0.30 _{- 0.08 }^{+ 0.17 }$ & $ 0.55_{- 0.26 }^{+ 1.36 }$ &  0.36 \\ [3pt] 

\hline
\end{tabular}
\label{tab:mamres}
\end{threeparttable}
\end{table}

\begin{table}
\centering
\caption{Parameter constraints for the composite cluster mass profiles. Columns represent the redshift range, virial radius \Rdyn,  scale radius $r_\mathrm{s}$, dynamical mass \Mdyn\ and concentration.  These results are obtained using Bayesian model averaging of the different anisotropy models.}
\begin{tabular}{cccccc}
\hline\\[-7pt]
Bin & Redshift &  \Rdyn & $r_\mathrm{s}$ & \Mdyn & $c_{200}$ \\ [2pt]
      & range & [Mpc] & [Mpc] & $[10^{14} M_{\odot}] $ &   \\[2pt]
\hline\\[-7pt]
1 & 0.26-0.38 & $ 1.81_{- 0.10}^{+ 0.11}$ &  $ 0.26_{- 0.10}^{+ 0.16}$ & $  9.44_{- 1.65}^{+ 1.70 } $  &  $5.4_{- 2.1}^{+ 2.6}$  \\[3pt]
2 & 0.39-0.44 &  $ 1.82_{- 0.09}^{+0.10}$ &  $ 0.36_{- 0.11}^{+ 0.17}$ & $  10.57_{-1.55}^{+1.93}$ &  $ 4.1_{- 1.3}^{+ 1.5}$   \\ [3pt]
3 & 0.46-0.56 & $ 1.56_{-0.06}^{+0.11}$  & $  0.31_{- 0.13}^{+ 0.15}$ & $ 7.42_{-0.92}^{+1.58}$  &  $ 4.4_{- 1.8}^{+ 1.7}$   \\   [3pt]
4 & 0.56-0.71   & $  1.51_{- 0.06}^{+ 0.05}$  &  $  0.25_{- 0.08}^{+ 0.11}$ & $ 7.31_{- 0.62}^{+1.13}$ & $ 5.2_{- 1.9}^{+1.9}$  \\ [3pt]
5 & 0.71-1.32 &  $ 1.28_{- 0.06}^{+ 0.06}$ & $ 0.24_{- 0.10}^{+ 0.11}$ &  $ 6.20_{- 0.88}^{+0.85}$  &  $4.2_{-1.6}^{+2.0}$  \\ [3pt]
- & 0.26-1.32 &  $1.62_{- 0.05}^{+ 0.03}$ & $ 0.29_{- 0.07}^{+ 0.06}$  & $ 8.71_{- 0.80}^{+ 0.52}$ & $ 5.1_{-1.0}^{+ 1.1}$   \\ [3pt]
\hline
\end{tabular}
\label{tab:results}
\end{table}

\section{Results}
\label{sec:results}

In this section we present the results of our dynamical analysis.   In the first subsection we explore the mass profiles and present several crucial pieces of information that are needed to understand the results in the following sections.  The following subsections present our measurements of the anisotropy profiles, the pseudo phase-space density profiles, and the dynamical mass constraints.  The final subsection examines the impact on our analysis of ongoing mergers.

\subsection{Testing Mass and Anisotropy Profiles}

As a first step, we use MAMPOSSt to analyze the composite cluster constructed using the full dynamical dataset on all mass and anisotropy models described in Section~ \ref{sec:MAMPOSSt}. Our goal is to determine which mass models are appropriate for this study. 
For each mass model, we explore all the different anisotropy models.

\subsubsection{Comparison of different mass profiles}

To quantitatively differentiate among the mass models (NFW, Einasto, Burkert, Hernquist and SIS), we compare the likelihood of the data being consistent with the model, employing the so-called Bayes factor.  This factor is the marginalized likelihood of the model \citep[see][and references therein]{Hoeting99bayesianmodel}. It is computed by averaging the likelihood in a specific model $M_{j}$ over the available prior range $P(\theta_{j}\,| M_{j})  $, reading
\begin{equation}
\mathcal{L}(D \,| M_{j}) = \int { \mathcal{L}(D\,| \theta_{j}, M_{j}) P(\theta_{j}\,| M_{j})  \text{d} \theta_{j}  } ,  
\end{equation}
where  $\mathcal{L}(D\,| \theta_{j}, M_{j})$ is the likelihood of the data $D$ given the model parameters $ \theta_{j}$.

Considering $J$ models $M_{1}$, ..., $M_{J}$, we define the Bayes factor $B_{j,max}$ of each model $j$ by normalizing by the most probable model, yielding
\begin{equation}
B_{j,max} = {\mathcal{L}(D \,| M_{j}) \over \mathcal{L}(D \,| M_{\text{max}})},
\label{eq:bayesfactor}
\end{equation}
where  $M_{\text{max}}$ indicates the model with the highest marginalized likelihood. The averaged posterior distribution on the parameters common to all models is then simply given by the weighted average of the posterior distributions of each model, with the Bayes factor as weight. Models can also be rejected using their Bayes factors.  According to the Jeffreys scale \citep{jeffreys1998}, $M_{j}$ is considered decisively rejectable if  $B_{j,\text{max}}<0.01$. 

Table~\ref{tab:massres} contains the measured values.  One can see that for each anisotropy model there is one preferred model (Bayes factor 1.0), but that the likelihood ratios for all but one of the mass profile models are on the order of 1.   Note that for the SIS profile we can only consider three of the five anisotropy models, because the T and O models need the value of the scale radius in the density profile, and that is not uniquely defined for the SIS.  Indeed, with the exception of the SIS model, we cannot reject any of the mass models we consider here. 

We note that our choice of parameter priors for these analyses does not affect the calculation of the Bayes factor, because these priors are set to be flat with allowed ranges that extend beyond the preferred range of each parameter. 

Taking guidance from both theoretical expectation and observational results, we select the NFW model as a good description of the mass profile. As mentioned in Section~\ref{sec:introduction}, in fact, cosmological simulations produce DM halos with mass profiles well described by an NFW profile. 
This result is in good agreement with a variety of observational analyses using both dynamics and weak lensing \citep{1997Carlberg, vanderMarel2000, 2003Biviano, Katgert2004, 2011Umetsu}, even though some results have preferred different models \citep{2006Merritt, 2010Navarro, 2014Dutton, 2015vdb, 2017Sereno}.

\subsubsection{Bayesian model averaging with anisotropy models}

In contrast to the mass profiles, the literature does not provide us with strong predictions for the radial form of the velocity anisotropy or $\beta$ profile. 
In Table~\ref{tab:mamres} we list-- for each of the composite clusters as defined in Table~\ref{tab:bins}-- the results of the MAMPOSSt analysis, with the anisotropy parameter $\theta_\beta$ being $\beta_{\text{C}}$ for the C model, $\beta_{\infty}$ for the T and O models and $r_{\beta}$ for the M\L\ and OM models. The errors on each of the parameters listed in the table are obtained by a marginalization procedure, i.e. by integrating the probabilities $p(r_{200}, r_{\text{s}}, \beta)$ provided by MAMPOSSt over the remaining two free parameters.  In addition to the anisotropy parameters, Table~\ref{tab:mamres} contains the dynamical constraints on the composite cluster virial radius \Rdyn\ and the Bayesian weight described above. 

Because we cannot strongly reject any of the models, we combine the results from the different anisotropy profiles by performing a Bayesian model averaging, weighting every model by its Bayes factor and combining statistics from the different $\beta$ models.   This approach was first proposed by \citet{2012aVazquez}, and has subsequently been used in multiple analyses aside from our own \citep{2012bVazquez,2014Aslanyan,2016Hee,2016PlanckXX}.  For a proof that the weighted and coadded probability distribution functions consistute a proper probability distribution function, we direct the reader to examine section 8.2, equation (69) of the last reference. The results of this analysis are shown in Table~\ref{tab:results} where we present the virial radius \Rdyn, the NFW scale radius $r_\mathrm{s}$, the virial mass \Mdyn, and the concentration $c_{200}$.  The 1$\sigma$ parameter uncertainties are computed through a marginalization procedure, as before. 

\subsubsection{Impact of Mass Priors on Composite Clusters}
As mentioned in Section~\ref{sec:SPTmasses}, when creating the composite clusters, we perform a rescaling of the observables (the galaxy projected cluster distances R and rest frame velocities $v_{\rm{rf}}$) using two different initial mass estimates, \Mszsigma\ and \MszPlanck\  (available for each individual cluster). The dynamical mass constraints that result from the composite clusters created using these two different scalings do not differ significantly,
 demonstrating the stability of our analysis to underlying systematic mass uncertainties in the initial mass estimates used to construct the composite clusters. We return to this point in Section~\ref{sec:compositemasses} below, where we show the dynamical masses obtained in the two cases (see Table~\ref{tab:masscomp}).  
 
 Moreover, the masses derived through the full dynamical analysis are in good agreement with the mean SPT based masses \Mszsigma\ listed in Table~\ref{tab:bins}. Regarding the precision of the constraints, it is clear that a composite cluster with $\sim$600 cluster members allows one to determine the dynamical mass with what is effectively a $\sim15$\% uncertainty ($\sim$8\% for the full dataset, using $\sim$3000 tracers). In contrast, the NFW scale radius and the corresponding concentration are only weakly constrained.

\subsubsection{Goodness of fit of dynamical models}
We have also examined whether the best fit models are an adequate description of the data.  To do this, we have created 1000 mock galaxy samples of similar size to the full galaxy sample by sampling the likelihood distribution in line of sight velocity and projected distance produced using the best fit model to the full galaxy sample.  The best fit model has an M\L\ anisotropy profile, and the Bayes factors of the other four models for the full galaxy sample are similar to those seen in Table~\ref{tab:mamres} for the redshift subsamples of the data, indicating that all five anisotropy models fit similarly well.  We then analyze each of these 1000 mock samples in the same way as the real data, examining the mean log(likelihood) per galaxy for the real data and the mocks.  The likelihood of the real data is somewhat lower than the typical likelihood of the mocks.  However, 6.1\%\ of the mock samples have even lower likelihood than the real data.  Thus, the dynamical models we are fitting are indeed an adequate description of the data.

\subsection{Velocity dispersion anisotropy profiles}
\label{sec:beta}

\begin{figure*}
\vskip-0.40in
\centering
\hbox to \hsize{\hskip-0.3in \includegraphics[width=200mm]{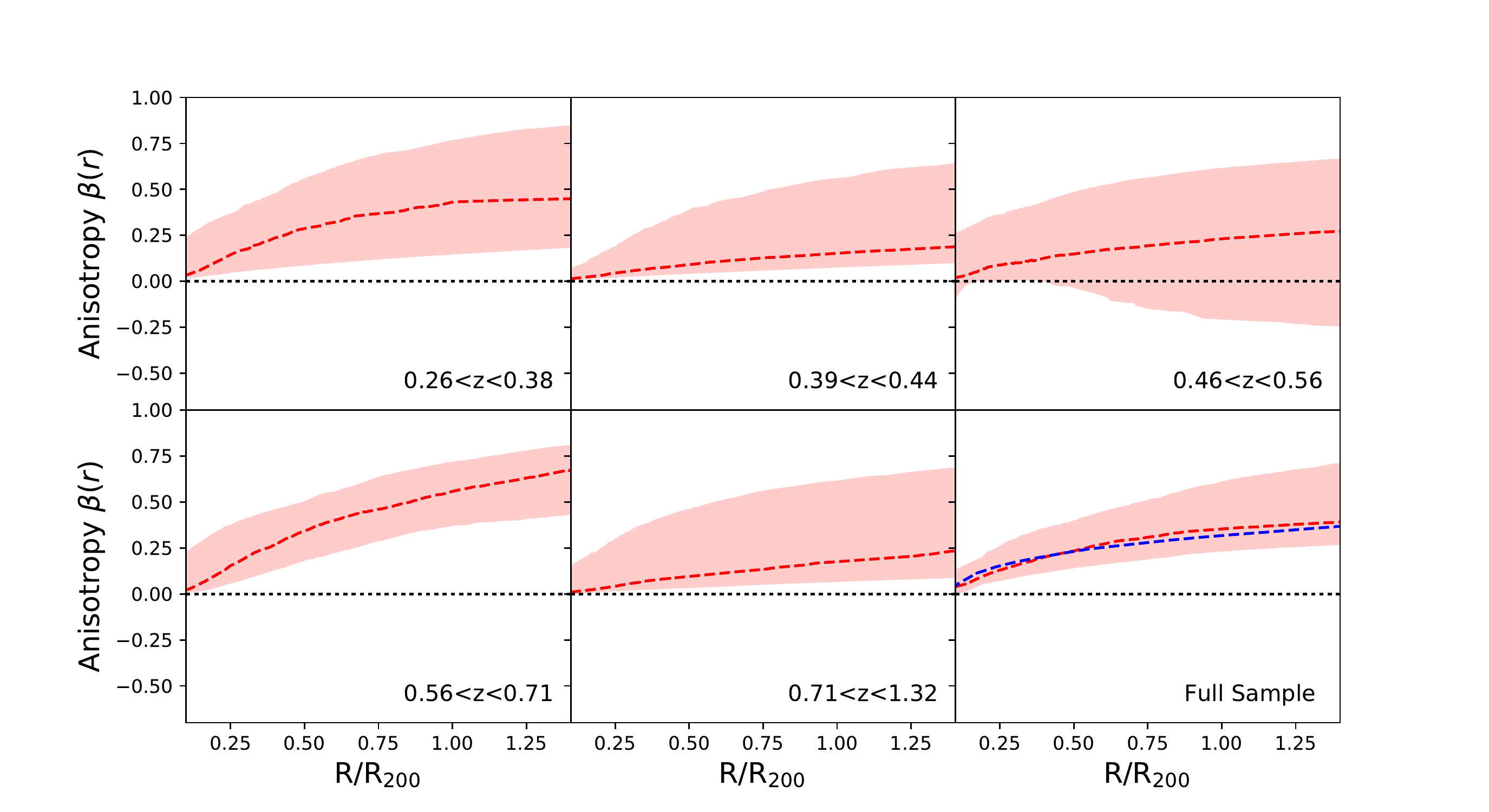} \hfil}
\vskip-0.15in
\caption{Velocity anisotropy profile $\beta(r)$ for each redshift bin. The red dashed line represents the profile obtained implementing a Bayesian model averaging, with the pink shaded region indicating the 1$\sigma$ confidence region around this solution.  There is no clear evidence of a redshift trend.  The blue line in the lower-right panel (full sample) shows the result obtained when adopting the best fit NFW model and the Jeans equation inversion to solve for the anisotropy profile.  This result is in good agreement with the model averaging result. Our analysis shows that passive galaxies preferentially move on nearly isotropic orbits close to the cluster center, and on increasingly radial orbits as one moves to the virial radius.}
\label{fig:anisotropy-redshift}
\end{figure*}
\begin{figure*}
\vskip-0.40in
\centering
\hbox to \hsize{\hskip-0.3in \includegraphics[width=200mm]{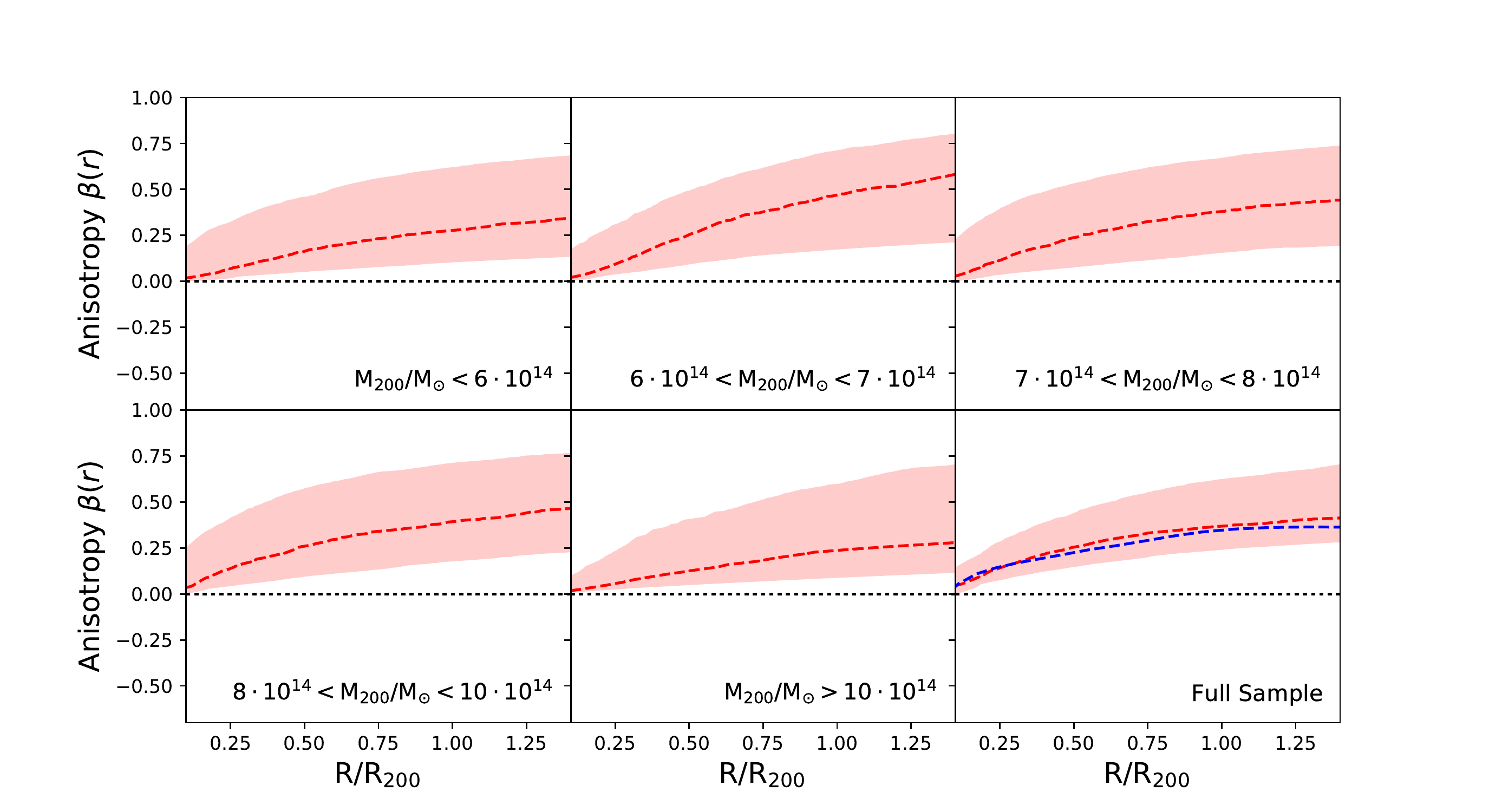} \hfil}
\vskip-0.15in
\caption{Velocity anisotropy profile $\beta(r)$ for each mass bin. As in Fig. 4, the anisotropy-redshiftred dashed line represents the profile obtained implementing a Bayesian model averaging, with the pink shaded region indicating the 1$\sigma$ confidence region around this solution. The blue line in the lower-right panel (full sample) shows the result obtained when adopting the best fit NFW model and the Jeans equation inversion to solve for the anisotropy profile. There is no clear evidence of a mass trend.}
\label{fig:anisotropy-mass}
\end{figure*}

In the following we first present the results of our examination of mass and redshift trends (Section~\ref{sec:massredshifttrends}) and then discuss implications for a particular mass accretion model (Section~\ref{sec:twophasemodel}) and then finnally compare with previous studies of the velocity dispersion anisotropy profile (Section~\ref{sec:anisotropyprofileprevious}).

\subsubsection{Constraints on Redshift and Mass Trends}
\label{sec:massredshifttrends}

Fig.~\ref{fig:anisotropy-redshift} contains the measured anisotropy $\beta(r)$ profiles and their 1$\sigma$ confidence regions for the five composite clusters in different redshift ranges together with the results from the full sample (lower, right-most panel).  
These profiles are obtained by using the posterior distribution in the anisotropy parameter $\theta_\beta$ extracted from each of the five anisotropy models.   Specifically, for each of the models a large number of $\theta_\beta$ values are drawn, consistent with the posterior.  Each value corresponds to an anisotropy profile.  The number of $\theta_\beta$ values drawn for each model is weighted according to the Bayes factor.  The sum of all these anisotropy profiles provides a measure of the probability distribution in the anisotropy profile value at each radius.  The red line represents the median value of this distribution, while the shaded region is defined by the 16th and 84th percentiles of the distribution (1$\sigma$ confidence region). As noted in the previous section, the mass profile model in all cases is an NFW with concentration and mass free to vary.

Our analysis indicates that the orbits of passive, red galaxies are nearly isotropic close to the cluster center, and become increasingly radial going towards larger radii, reaching a radial anisotropy $\beta \simeq 0.15-0.6$ at  $R/r_{200} \simeq 1$.  There is no clear evidence for a redshift trend in the anisotropy profile of the passive galaxy population out to $z\approx1$. We have carried out a similar analysis of five composite clusters built from the same cluster sample divided into mass bins rather than redshift bins, and we find no evidence for trends with mass, either. We show this result in Fig.~\ref{fig:anisotropy-mass}. 

For this reason we analyze also the full sample, providing our best available constraints.  The orbital anisotropy varies from values consistent with zero in the cluster core to a value $0.4\pm0.15$ at the virial radius.  For reference, anisotropy values of 0.4 correspond to tangential components of the velocity dispersion ellipsoid having amplitudes that are only 60~percent as large as the radial component. For the full sample we show (blue dashed line) also the anisotropy profile recovered using the best fit NFW parameters and using the Jeans equation to solve for the velocity dispersion anisotropy profile \citep{1982Binney,1990Solanes,2013Biviano}. This result is in good agreement with the solution recovered using the Bayesian model averaging over the five adopted anisotropy profiles.  

\subsubsection{Comparison with Two-Phase Accretion Model}
\label{sec:twophasemodel}

The behavior of the anisotropy profile is consistent with the theoretical model discussed in \citet{2009Lapi}, according to which the growth of structure proceeds in two phases: an early, fast accretion phase during which the cluster undergoes major merging events, and a second slower accretion phase involving minor mergers and smooth accretion \citep[see, e.g.,][]{1986White, 2003Zhao, 2007Diemand}.  \citet{1967LyndenBell} discusses how a dynamical system rapidly relaxes from a chaotic initial state to a quasi-equilibrium. The first stage of fast accretion provokes rapid changes in the cluster gravitational potential, inducing the collisionless components of the cluster to undergo violent relaxation, resulting in orbits that are more isotropic.
Galaxies accreted by the cluster during the fast accretion phase would then be expected to exhibit approximately isotropic orbits, while galaxies accreted during the second phase would maintain their preferentially radial orbitals over longer timescales.  Given that as the cluster accretes its mass and virial radius also grow, a two phase scenario like this would tend to lead to anisotropy profiles that are isotropic in the core and become more radial at larger radius.  

However, one might also expect to see a time or redshift variation of the anisotropy profile, with typical galaxy orbits in high redshift clusters showing less of a tendency for radial orbits near the virial radius.  The fact that our analysis shows no strong redshift trend in the anisotropy profiles is an indication that, in massive galaxy clusters, the passive galaxy population orbits are not changing significantly with cosmic time since $z \approx 1$.  This suggests that the merging and relaxation processes responsible for the anisotropy profiles are underway at all cosmic epochs probed here.  Other indicators of cluster merging have shown similar results; namely, the fraction of systems with disturbed X-ray morphologies \citep[typically measured using centroid variations or ellipticities;][]{mohr93} does not change significantly with redshift in samples of homogeneously SZE selected cluster samples \citep{2017Nurgaliev, 2017McDonald}.  This may suggest that the fast accretion phase of such a two phase model could be very short and happening primarily at redshifts above those probed by our sample.  
Clearly, a detailed examination of cosmological $N$-body simulations with sufficient volume to contain rare, massive clusters and sufficient mass resolution to ensure the survival of galaxy scale subhalos after accretion into the cluster is warranted.  Moreover, a dynamical analysis like the one we have carried out that focuses on systems at redshift $z>1$ would enable a more sensitive probe for time variation in the growth of structure.

\subsubsection{Comparison with Previous Results}
\label{sec:anisotropyprofileprevious}

Previous studies of the velocity dispersion anisotropy conducted on passive cluster members at intermediate to high redshifts present hints of an anisotropy profile that is nearly isotropic close to the cluster center, and increasingly radial at larger radii. 
\citet{2009Biviano}, stacking 19 clusters between redshift $\approx 0.4-0.8$ with a mean mass of $\approx 3 \times 10^{14} M_{\odot}$, suggest radially anisotropic orbits. \citet{2016Biviano} reach the same conclusion when analysing a stacked sample of 10 clusters at $0.87 < z < 1.34$. Studying a single cluster, \citet{2016Annunziatella} show that the same trend is found for galaxies characterized by a stellar mass $M_{\star} > 10^{10} M_{\odot}$, while lower-mass galaxies move on more tangential orbits, avoiding small pericenters, presumably because those that cross the cluster center are more likelly to have been tidally destroyed. These results are consistent with ours.  Our result, obtained through the analysis of a large sample of passive galaxies within a homogeneously selected sample of massive clusters over a wide redshift range and with low scatter mass estimates, allows us to cleanly probe for redshift and mass trends in the velocity dispersion anisotropy profile.

Some published analyses carried out at lower redshifts ($z \lesssim 0.1$) than our sample show similar results.  \citet{2010Wojtak} analyzed a sample of 41 nearby relaxed clusters, finding that galaxy orbits are isotropic at the cluster centers and more radial at the cluster virial radius. A similar result is obtained by \citet{2009Lemze} and \citet{2017Aguerri}. However, other analyses show that the orbits of passive galaxies at these redshifts are more isotropic at all radii \citep{2004Biviano, Katgert2004, 2009Biviano, 2014Munari}, hinting at a possible change in galaxy orbits over time due to processes such as violent relaxation, dynamical friction, and radial orbital instability \citep{2008Bellovary}. 
At present, results from numerical simulations predict a range of behavior \citep{2011Wetzel, 2012Iannuzzi, 2013Munari}, so further study is definitely needed. Extending our own observational analysis towards lower redshifts could also help clarify this picture.

\subsection{Pseudo phase-space density profiles}
\label{Q(r)}

\begin{figure*}
\centering 
\vskip-0.40in
\hbox to \hsize{\hskip-0.30in\includegraphics[width=200mm]{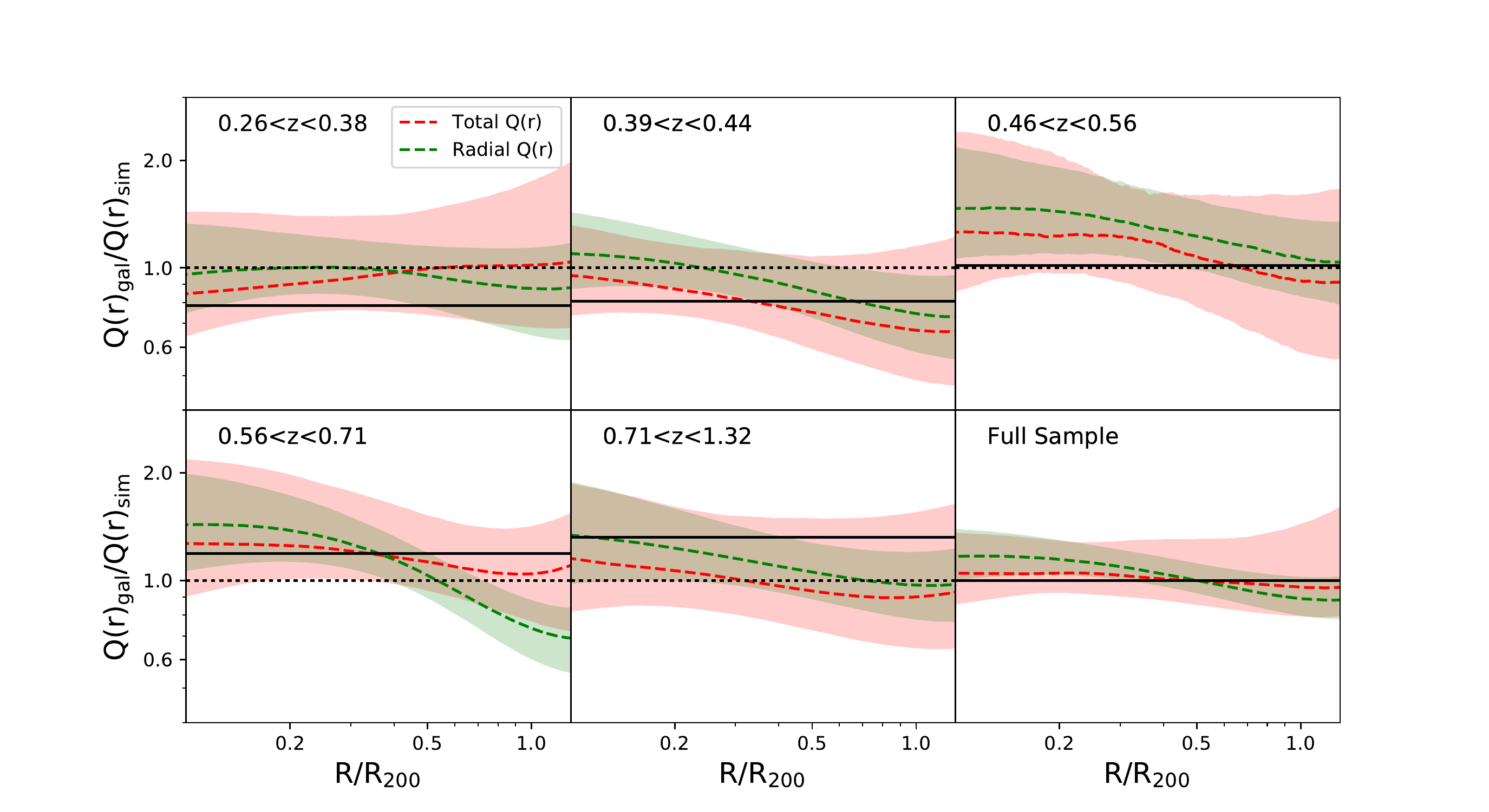} \hfil}
\vskip-0.18in
\caption{The ratios of the passive galaxy pseudo-phase-space density (PPSD) profiles $Q_\textrm{gal}(r)$ to that of dark matter particles in simulations $Q_\textrm{sim}(r)$  \citep{2005Dehnen} are shown in green for the total PPSD $\rho_\textrm{gal} / \sigma^{3}$ and in red for the radial PPSD $\rho_\textrm{gal} / \sigma_\mathrm{r}^{3}$ for five redshift bins and the full sample.  Shading corresponds to the 1$\sigma$ confidence region.  The normalization of each $Q(r)$ profile (both of the galaxies and of the simulations) is fixed to that of the $Q_\textrm{gal}(r)$ of the full sample, which is determined by fitting to the data. Thus, the deviations from $Q_\textrm{gal}(r)/Q_\textrm{sim}(r) =1$ of the full sample are only driven by the deviations of the $Q_\textrm{gal}(r)$ slope from the  $Q_\textrm{sim}(r)$ one, but in the other samples they are also driven by real normalization differences. The solid black line marks the expected amplitude in each redshift bin under the assumption of cluster self-similar redshift evolution.   Both passive galaxy PPSD profiles are in reasonably good agreement with $N$-body simulations, and the data are roughly consistent with self-similar scaling with mass and redshift.}
\label{fig:Q}
\end{figure*}

The determination of the anisotropy profile $\beta(r)$ allows us to investigate the behavior of the Pseudo Phase-Space Density profile $Q(r)$ introduced in Section~\ref{sec:introduction}. According to numerical simulations of virialized halos \citep{2001Taylor, 2005Dehnen, 2009Lapi}, there is a scaling between the density $\rho$ and the velocity dispersion, best appreciated by considering the quantity $Q(r) = \rho/\sigma^{3}$. The profile of this quantity is found to follow a universal power-law of fixed slope, $\propto r^{-1.875}$. Remarkably, this is the same power-law predicted by the similarity solution of \citet{1985Bertschinger} for secondary infall and accretion onto an initially overdense perturbation in an Einstein-de Sitter universe. In that work the authors found that the relaxation process is self-similar, meaning that each new shell falling in is virialized and adds a constant contribution to the resulting power-law density profile.  

Because the velocity dispersion profile is constrained by the galaxies, we derive the PPSD profile using the number density profile of galaxies $\rho_\mathrm{gal}$ instead of the mass density profile.  For each stacked cluster, we fix the Halo Occupation Number $N_{200}$ of red galaxies to that found by H17 to find the central density $\rho_{0}$ of the NFW $\rho_\mathrm{gal}$ profile, such that 
\begin{equation}
\rho_{0}=\dfrac{N_{200}}{4 \pi \int_{0}^{R_{200}} \frac{r^{2}}{r/r_{s}(1+r/r_{s})^{2}}dr }.
\end{equation}
 
We present the results obtained investigating both the total PPSD profile $Q_\mathrm{gal}(r) \equiv \rho_\mathrm{gal} / \sigma^{3}$ and the radial PPSD profile $Q_\mathrm{gal,r}(r) \equiv \rho_\mathrm{gal} / \sigma_{r}^{3}$, where $\rho_\mathrm{gal}$ is obtained as described above and $\sigma_{r} (r)$ is recovered using the following equation \citep{1994VdM, 2005Mamon, 2013MAMPOSSt} 
\begin{equation}
\sigma_{r}^{2}(r) = \frac{1}{\nu (r)} \int_{r}^{\infty} { \exp \left[ 2 \int_{r}^{s} { \beta(t) \frac{dt}{t} } \right] \nu(s) \frac{G M(s)}{s^{2}}  ds    }, 
\end{equation}
evaluated over an adequate grid of $r$, and $\sigma(r)$ is given by
\begin{equation}
\sigma (r) = \sqrt{3 - 2 \,\, \beta(r)}  \,\,  \sigma_{r}(r)  . 
\end{equation}

Fig.~\ref{fig:Q} shows the ratio between the derived galaxy PPSD profiles and the fixed-slope best-fit relations $Q(r) \propto r^{-1.84}$ and $Q_{r}(r) \propto r^{-1.92}$, where the slopes are those obtained by \citet{2005Dehnen} by studying DM particles in numerically simulated halos, while the normalizations are fitted to the data of the full sample. A similar value for the slope of $Q(r)$ is found by \citet{2007Faltenbacher}, a work based on numerical simulations and focused on the gas and DM entropy profiles in galaxy clusters. The similarity between radial profiles of the passive galaxy PPSD and simulated PPSD profiles is an indication that the passive galaxy radial distribution and kinematics are similar to those of the dark matter particles in those simulations.  

We emphasize here that it is not possible for us to predict the PPSD profile for the underlying dark matter in our sample, because in general the blue or EL  galaxy population exhibits different kinematical properties than those of the passive, red or nEL galaxies.  One might expect the full dark matter kinematics to exhibit a mix of the properties of the different galaxy populations.  
We note that because the measured concentration of the passive population (H17) is similar to the derived concentration of the mass profiles from our analysis, that if we replace the galaxy density profile with our derived mass density profile, the PPSD profiles have a very similar radial behavior. 

In Fig.~\ref{fig:Q} we also show (solid black lines) the expected amplitude if there is self-similar behaviour of passive galaxy profiles and kinematics.  That is, we have $Q(r) \equiv \rho/\sigma^{3}   \propto  E(z)^{2}/[M(r)E(z)] = E(z)/M(r)$, where  $H(z)=H_0E(z)$ is the Hubble parameter. The passive galaxy PPSD amplitudes are in reasonably good agreement with the amplitudes expected under self-similarity, providing some additional evidence that this population is approximately self-similar.

In the previous sub-section, we mentioned that the anisotropy profiles could be the result of violent relaxation. This process, driven by gravity alone, would also tend to create a scale-invariant phase-space density.  Relaxation into dynamical equilibrium would then lead the slope of the PPSD profile to approach a critical value, resulting in the particular form of the density profile for DM particles within simulated halos. An anisotropy profile isotropic in the inner regions and increasingly radial at larger radii gives the pseudo phase space density profile slope observed in numerical simulations. The agreement then suggests that the passive galaxies we analyze here have reached a similar level of dynamical equilibrium as the dark matter particles in the simulations, and that this is true for all redshifts up to $z\simeq1$. 

It is difficult to compare our results directly to previous studies, because most of those studies have tended to focus on $Q(r)$ and $Q_\mathrm{r}(r)$ using the inferred total matter density from their dynamical analyses rather than the galaxy density profile.  Such an approach produces a mixed PPSD profile that contains both total matter and galaxy properties, because the dispersion profiles used are necessarily those of the galaxies rather than the dark matter.  This complicates the interpretation, because a mismatch between the observed and simulated PPSD profiles could be because the total matter density profile doesn't match the simulated dark matter profile or it could mean that the velocity dispersion profile of the dark matter and a particular galaxy population do not match.  In general, given that different galaxy populations tend to exhibit different kinematic properties (i.e. radial distributions and velocity dispersion and anisotropy profiles), one would not expect the mixed PPSD profiles defined in this way to agree for the different populations.

Indeed, \citet{2013Biviano} measured $Q(r)$ using the anisotropy profile of the galaxies and the NFW mass density profile inferred from their analysis, separately for star forming (SF) and passive galaxies in a single cluster. They found that their observed PPSD profile agreed with simulations only for the passive galaxy population.   They argue that passive members have undergone violent relaxation and have reached dynamical equilibrium, while SF members have not yet reached equilibrium.  But another possible interpretation would be that the Jeans analyses of both populations are equally accurate, but that the passive population exhibits a velocity dispersion and anisotropy profile more similar to that of the simulated dark matter.  A more recent study came to similar conclusions \citep{2014MunariErr}.

On the other hand, a recent study of the nearby cluster Abell~85 \citep{2017Aguerri} shows that $Q_{r}(r)$ follows the theoretical power-law form independent of the galaxy colour or luminosity, concluding that all the different families of galaxies under study reached a virialized state.  Given the discussion above, this agreement in the mixed PPSD profile derived from different galaxy populations also indicates that the different populations must have similar kinematics (i.e. velocity dispersion and anisotropy profiles).  In their study, they emphasize that the anisotropy profiles of the blue and red galaxies are different, which would make the agreement in $Q_\mathrm{r}(r)$ surprising.  However, Fig.~3 in their paper suggests that the anisotropy profiles have similar character (isotropic in the center, more radial at larger radius) and present evidence for inconsistency that is weak ($\le2\sigma$).

\begin{table*}[htb]
\caption{Comparisons of dynamical masses from composite clusters calculated using different initial masses.  From left to right the columns contain the redshift range of  the cluster sample, the derived dynamical mass given an initial SPT plus velocity dispersion mass, the derived dynamical mass given an SPT + Planck initial mass, and the ratios of these dynamical to initial masses in each case. Finally, we report the constraints on $\eta$ and $\eta'$ as described in Section~\ref{sec:masstension}.}
\centering
\begin{tabular}{cccccccc}
\hline\\[-7pt]
Bin &Redshift & $M_{200}^{\text{dyn}}\,|$ \Mszsigma\   &  $M_{200}^{\text{dyn}} \,| $ \MszPlanck\  & $\dfrac{M_{200}^{ \text{dyn}} \,| M_{200}^{\text{SZ}+\sigma}} {\left< M_{200}^{\text{SZ}+\sigma} \right>}$ & $\dfrac{M_{200}^{\text{dyn}} \,| M_{200}^{\text{SZ}+\text{Planck}}} {\left< M_{200}^{\text{SZ}+\text{Planck}} \right>}$  & $\eta$ & $\eta'$ \\  [-2pt]
& range  &   $[10^{14} M_{\odot}] $ & $[10^{14} M_{\odot}] $   & & & \\  [2pt]
\hline\\  [-7pt]
1 & 0.26-0.38  & $9.44_{-1.65}^{+1.70} $ & $9.29_{- 1.45}^{+ 1.96 }$ & $1.02_{-0.18}^{+ 0.18} $ & $0.86_{-0.15}^{+0.15} $  & $0.77_{-0.13}^{+0.28} $ & $1.21_{-0.36}^{+0.17} $\\[2pt]
2 & 0.39-0.44   & $  10.57_{-1.55}^{+ 1.93}$ &  $  10.41_{-1.19}^{+ 1.88}$ & $1.11_{-0.16}^{+ 0.20}$ &$ 0.91_{- 0.13}^{+0.17} $ &$ 0.92_{- 0.20}^{+0.17} $ & $1.24_{-0.27}^{+0.21} $\\[2pt]
3 & 0.46-0.56   & $  7.42_{- 0.92}^{+ 1.58}$  & $  7.14_{-  0.88}^{+1.49}$  & $ 0.93_{-0.11}^{+ 0.20} $ & $0.74_{-0.09}^{+0.16} $ & $0.74_{-0.09}^{+0.15} $  & $0.97_{-0.12}^{+0.23} $\\[2pt]
4 & 0.56-0.71   & $ 7.31_{- 0.62}^{+ 1.13}$   & $ 7.36_{-  0.78}^{+ 1.01}$  & $1.00_{- 0.08}^{+ 0.15} $ &$ 0.73_{-0.06}^{+0.11} $ &$ 0.79_{-0.10}^{+0.10} $ & $1.10_{-0.13}^{+0.13} $ \\[2pt]
5 & 0.71-1.32 & $ 6.20_{-  0.88}^{+ 0.85}$  & $ 5.95_{- 0.75}^{+0.83}$ & $0.82_{- 0.12}^{+0.11}$ &$ 0.55_{-0.08}^{+ 0.07} $ &$ 0.63_{-0.08}^{+ 0.13} $  & $0.90_{-0.11}^{+0.18} $ \\[2pt]
- & 0.26-1.32 &  $ 8.71_{- 0.80}^{+ 0.52}$ & $ 8.50_{-0.67}^{+0.59}$ & $1.05_{-  0.10}^{+ 0.06}$ & $0.81_{-0.07}^{+0.05} $  & $0.81_{-0.06}^{+0.06}$ & $1.14_{-0.07}^{+0.06} $ \\[2pt]

\hline
\end{tabular}
\label{tab:masscomp}
\end{table*}

\subsection{Dynamical mass constraints}
\label{sec:masses}

We use the dynamical information to study the masses of these clusters in two different ways.  In Section~\ref{sec:compositemasses}, we analyze the composite clusters and the consistency of the dynamical masses when using different initial masses to scale the galaxy observables.  In Section~\ref{sec:masstension}, we examine the differences between the dynamical masses and the SPT+Planck masses $M_{200}^{\text{SZ}+\text{Planck}}$ using a different approach where the scaling values for each cluster $r_{200}$ and $v_{200}$ are altered self-consistently in each iteration of the Markov chain to reflect the SPT+Planck masses scaled by a free parameter $\eta$, defined in equation~(\ref{eq:eta}).

\subsubsection{Mass constraints on the composite clusters}
\label{sec:compositemasses}

\begin{figure*}
\vskip-0.40in
\centering 
\hbox to \hsize{\hskip-0.3in\includegraphics[width=200mm]{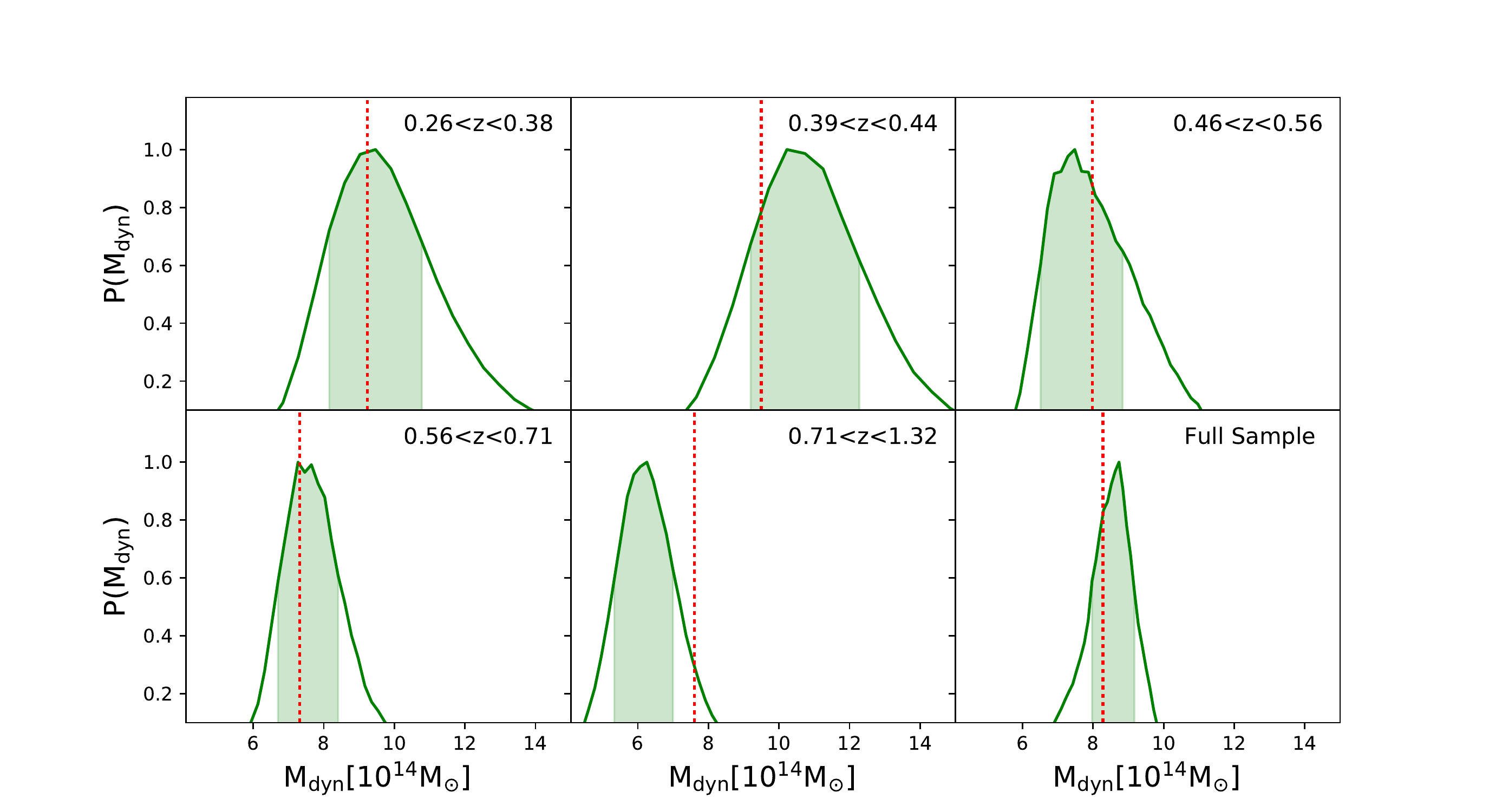}\hfil}
\vskip-0.15in
\caption{Marginalized distribution of the dynamical masses. Each panel corresponds to a different redshift range, and the final panel shows the results of the analysis of the full sample.  In green we highlight the 1$\sigma$ confidence region. The red line represents the mean SPT+$\sigma$ mass $\left<M_{200}^{\text{SZ}+\sigma}\right>$ for the clusters in the bin, weighted by the number of member galaxies in the individual clusters. There is a good agreement between the dynamical masses and the originally inferred SPT masses in all cases (see Table~\ref{tab:masscomp}).}
\label{fig:Mvd}
\end{figure*}

Constructing the composite clusters (described fully in Section~\ref{sec:stack}) requires a scaling of our galaxy observables, $v_{\text{rf}}$ and $R$, by estimates of the virial radius and velocity $r_{200}$ and $v_{200}$ respectively, and therefore requires an initial mass estimate.  As described above, this is a potential problem, because there is currently a $\sim$25~percent shift between the SPT cluster masses derived using cluster counts and velocity dispersion information, and the masses derived using the cluster counts and external cosmological priors from Planck \citep{bocquet15,2016deHaan}.  Thus, we examine the dynamical mass constraints in each case:  (1) those derived using initial masses derived from the cluster counts and velocity dispersion measurements $P(M_{200}^{\text{dyn}}\,| M_{200}^{\text{SZ}+\sigma})$ and (2) those derived using initial masses from the cluster counts and external cosmological constraints from Planck $P(M_{200}^{\text{dyn}}\,|M_{200}^{\text{SZ}+\text{Planck}})$.

In Fig.~\ref{fig:Mvd} we display the marginalized distribution of the dynamical masses obtained with our analysis where the $M_{200}^{\text{SZ}+\sigma}$ masses were adopted for the initial scaling.  The green regions mark the 1$\sigma$ confidence regions, and the red lines represent the mean initial masses derived from the cluster counts and velocity dispersion measurements $\left<M_{200}^{\text{SZ}+\sigma}\right>$, where the masses from each cluster are weighted by the number of galaxy velocities available for that cluster. There is good agreement between the dynamical masses and the initial masses in all redshift bins and also for the full sample (lower, right-most panel).  The second column of Table~\ref{tab:masscomp} contains the measurement results and uncertainties for each subset.  Characteristic dynamical mass uncertainties are at the $\sim$15~percent level for individual subsamples and at the $\sim$8~percent level for the full sample.  These are quite encouraging mass constraints, given that they are marginalized over the velocity dispersion anisotropy profile uncertainties.

We find that by fixing the concentration and radial anisotropy parameters to their best fit values when fitting for mass in a composite cluster with 600 tracers, the resulting mass uncertainty is not significantly impacted.  This suggests that our uncertainties in the individual redshift bins are not dominated by the freedom in mass and anisotropy profiles.  A similar test on the composite cluster built from the full dynamical sample leads to a $\sim$5\% mass uncertainty, which is comparable to what we find when using a single anisotropy model, before performing the Bayesian model averaging.  This is an interesting result when taken together with the discussion of velocity dispersion based mass estimates in \citet{2016Sifon}, where the scaling presented suggests that with samples of 600 dynamical tracers mass estimates should be closer to $\sim$7\% accurate rather than the 15\% we recover.  Further examination of the assumptions built into the dynamical mass measurements using velocity dispersions and full Jean analysis modeling is warranted and is planned for a future analysis.

As discussed in Section~\ref{sec:MAMPOSSt}, current estimates from studies of clusters in numerical simulations indicate there are remaining systematic uncertainties associated with MAMPOSSt analysis at the 10\% level.

To test the stability of the recovered dynamical masses to the initial input masses used for scaling, we perform the same analysis using the SPT+Planck cluster masses.  The third column of Table~\ref{tab:masscomp} shows these results.  These dynamical masses with the different initial masses are quite close to the values derived with the other set of initial masses.  This shows that there is no strong dependence of the dynamical mass on the initial mass.  This is because any change in the masses used for rescaling the cluster observables during stacking will impact, on average, the individual cluster masses and the final mean mass in the bin in a similar manner.  The overall scale of the dynamical data in projected radius and LOS velocity remains approximately invariant.

As columns four and five of Table~\ref{tab:masscomp} make clear, the dynamical masses, while being in good agreement with the cluster counts plus velocity dispersion masses $M_{200}^{\text{SZ}+\sigma}$, exhibit some discordance with the cluster counts plus external cosmological constraint masses $M_{200}^{\text{SZ}+\text{Planck}}$.  While the three lowest redshift bins show no significant disagreement, the upper two redshift bins show masses that are only 73~percent and 55~percent as large as the SPT+Planck masses (offsets that are statistically significant at the 2.5$\sigma$ and 6.5$\sigma$ levels, respectively).  The full sample has a dynamical mass that is only 80~percent of the SPT+Planck masses, a difference that is significant at the 3.8$\sigma$ level (statistical only).  The direction and scale of this mass shift is similar to that highlighted already in \citet{bocquet15}.  However, with our analysis we are able to show that this discrepancy seems to grow with redshift.

\begin{figure*}
\vskip-0.40in
\centering 
\hbox to \hsize{\hskip-0.3in\includegraphics[width=200mm]{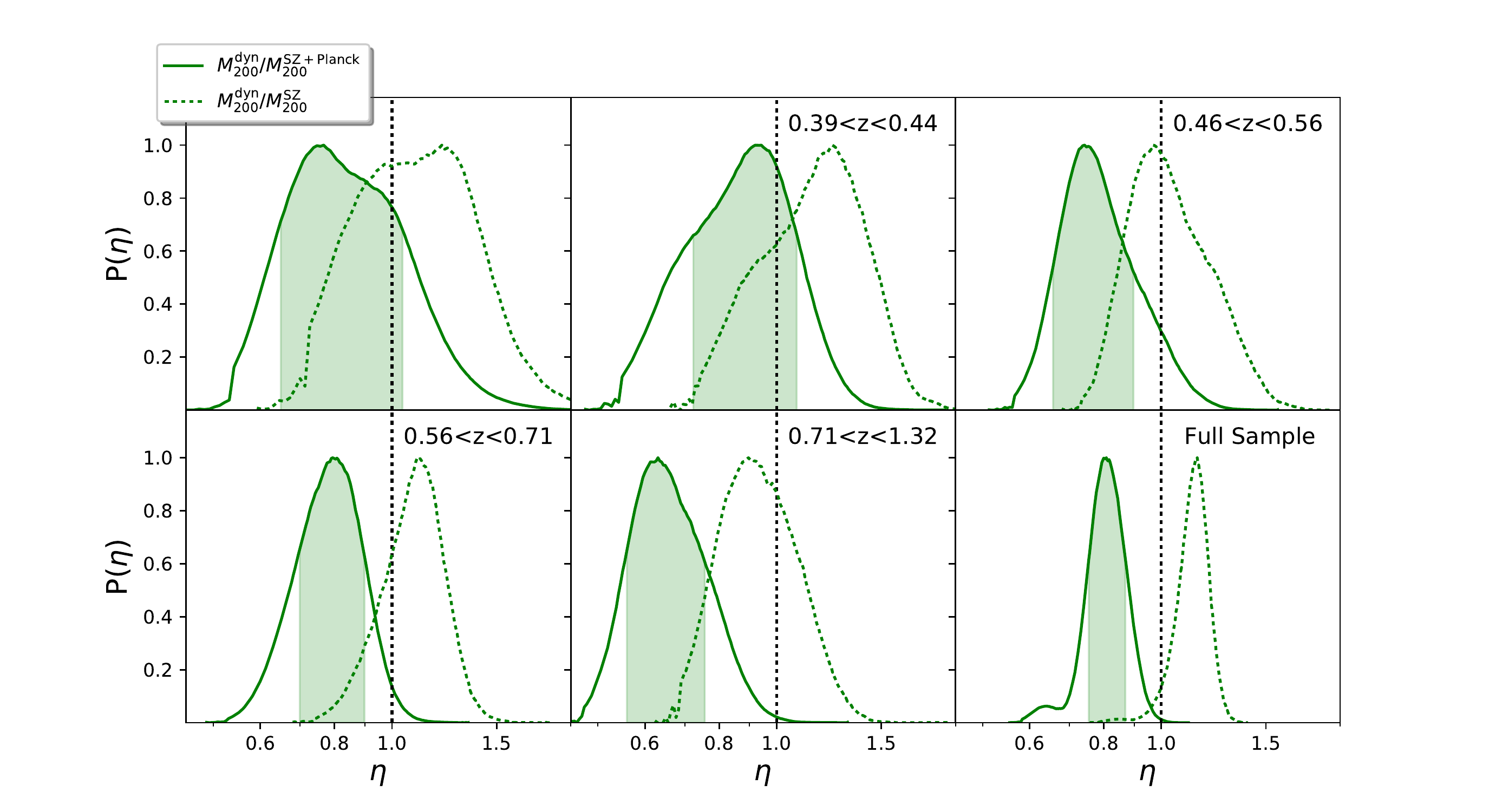} \hfil}
\vskip-0.15in
\caption{Posterior distribution of  $\eta =  M_{200}^{\text{dyn}}/M_{200}^{\text{SZ}+\text{Planck}}$ (solid green lines) arising from a dynamical analysis of each cluster subsample.  The green region shows the 1$\sigma$ region, while the vertical dotted black line marks the value $\eta=1$.  These distributions show that in the two high redshift bins and for the full sample there is disagreement between the dynamical masses and the SPT+Planck calibrated cluster masses.  For the full sample the discrepancy is 1.9$\sigma$ significant when including estimates of the systematic uncertainties.  In the highest redshift bin, the discrepancy is 2.6$\sigma$.  The dashed green lines show the estimated posterior for $\eta'=M_{200}^{\text{dyn}}/M_{200}^{\text{SZ}}$, where $M_{200}^{\text{SZ}}$ represents the \citet{2016deHaan} masses calibrated using the SPT mass function and $Y_X$ measurements for many of the systems.  In contrast to the SPT+Planck masses, these masses are in good agreement ($1\sigma$ offset) with the dynamical masses.}
\label{fig:eta}
\end{figure*}

\subsubsection{Comparison with SZE based masses}
\label{sec:masstension}

To examine this discrepancy more carefully, we use the dynamical analysis to examine the masses of these clusters and compare them to the masses derived separately from the SPT cluster counts in combination with external cosmological constraints from the Planck CMB anisotropy.
Rather than using the composite clusters, we analyze individual clusters, combining the likelihoods from each cluster and exploring constraints on an overall mass scaling parameter $\eta$, that is defined as
\begin{equation}
\eta =  \frac{M_{200}^{\text{dyn}}}{M_{200}^{\text{SZ}+\text{Planck}}}.
\label{eq:eta}
\end{equation}
We do this by running MAMPOSSt for each individual cluster in our sample. We calculate the posterior distribution of $\eta$ by using a multimodal nested sampling algorithm, namely MultiNest \citep{2008Feroz, 2009Feroz, 2013Feroz}, which provides us with the evidence for each model, and allows us to perform a Bayesian model averaging over different subsets of clusters.

Fig.~\ref{fig:eta} contains a plot of the posterior distributions of $\eta$ from our analysis within each redshift bin and for the full sample.  Results are largely consistent with the results from the composite clusters.  Column six of Table~\ref{tab:masscomp} contains the best fit $\eta$ values and associated uncertainties.  The preferred value for the full sample is $\eta=0.81 \pm 0.06\pm0.08$.  The constraint is followed by a statistical uncertainty and then a systematic uncertainty.  

As already discussed in Section~\ref{sec:MAMPOSSt}, studies of dynamical tracers drawn from clusters in structure formation simulations indicate that MAMPOSSt derived dynamical masses has systematic uncertainties of  $\approx10\%$ \citep[see][]{2013MAMPOSSt}.  This number comes from an analysis of tracers lying within a sphere of radius $r_{100}$ that are then used to estimate the virial radius $r_{200}$.  For the systematic uncertainty presented above, we have therefore adopted as a Gaussian with $\sigma=10$\% centered at no bias.

If one combines the statistical and systematic uncertainty in quadrature, the implication would be a difference at the $1.9\sigma$ level.  As mentioned already, this tendency for the dynamical masses to be lower than those masses derived from the cluster counts in combination with external cosmological constraints is consistent with the tendencies seen previously \citep{bocquet15} using simply dispersions and the cluster mass function in \citep[see also][]{2016Rines,2016Sifon}  More recent weak lensing analyses also support the lower mass scale of SPT clusters \citep{dietrich17,stern18}. 

To emphasize, Fig.~\ref{fig:eta} also shows (dotted line) the distribution of 
\begin{equation} 
\eta'= \frac{M_{200}^{\text{dyn}}}{M_{200}^{\text{SZ}}},
\label{eq:etaprime}
\end{equation}
where $M_{200}^{\text{SZ}}$ are masses calibrated from a cosmological analysis carried out in \citet{2016deHaan} using the X-ray mass proxy $Y_{x}$ and the abundance of clusters as a function of redshift,  $Y_{x}$ and the SZE mass proxy $\xi$, without the inclusion of the external Planck cosmological constraints.  These results indicate that the dynamical masses are in good agreement with the $M_{200}^{\text{SZ}}$ masses at all redshifts and for the full sample.  The $\eta'$ distribution for the full sample prefers a value of $\eta'=1.14\pm0.07\pm0.11$, indicating no disagreement.

To summarize, our dynamical mass measurements, which are derived using only dynamical information and no information from the mass function or cluster counts, are in good agreement with masses derived using information from the the cluster counts together with additional information from either velocity dispersions or from X-ray $Y_X$ measurements that have been externally calibrated.  However, our mass measurements exhibit moderate disagreement with those masses obtained similarly but when also adopting external cosmological priors from Planck CMB anisotropy.  Progress in testing these two mass scales would require better control of the systematic uncertainties in the dynamical masses \citep{2013MAMPOSSt}.  The agreement between the dynamical and the Planck based masses is best at low redshift, with the dynamical masses preferring ever smaller $\eta$ with increasing redshift.  In the highest redshift bin ($0.71\le z\le1.32$) we measure $\eta=0.63^{+0.13}_{-0.08}\pm0.06$, discrepancy at the 3$\sigma$ level.

\begin{table}
\centering
\caption{Sensitivity of dynamical mass measurements to the dynamical state of clusters.  We compare the masses for those clusters exhibiting large X-ray surface brightness asymmetries (un-relaxed) and those exhibiting small asymmetries (relaxed).  Columns list the subsample, the mean redshift, concentration, dynamical mass given initial scaling using SPT dispersion based masses, and ratio of the dynamical mass to the SZE dispersion based mean mass for the subsample.}
\begin{tabular}{lcccc}
\hline
Subsample & $\langle z \rangle$ & $c_{200}$  & $M_{200}^{\text{dyn}}\,| M_{200}^{\text{SZ}+\sigma}$ & $\dfrac{M_{200}^{ \text{dyn}} \,| M_{200}^{\text{SZ}+\sigma}} {\left< M_{200}^{\text{SZ}+\sigma} \right>}$ \\ 
 & &   & $[10^{14} M_{\odot}] $ &    \\ \hline\\ 
relaxed &$0.57 $ & $6.4_{-  2.4}^{+ 1.4}$ & $  7.0_{-0.9}^{+ 0.8} $ & $0.8_{-0.1}^{+ 0.1}$ \\[3pt]
un-relaxed & $0.6$ & $4.5_{-1.1}^{+1.2}$ &  $ 9.5_{- 1.2}^{+1.0} $  & $1.0_{-0.1}^{+ 0.1}$ \\[3pt]
\hline
\end{tabular}
\label{tab:results xrays}
\end{table}

\subsection{Impact of disturbed clusters}

\begin{figure}
\vskip-0.10in
\centering 
\includegraphics[scale=0.50]{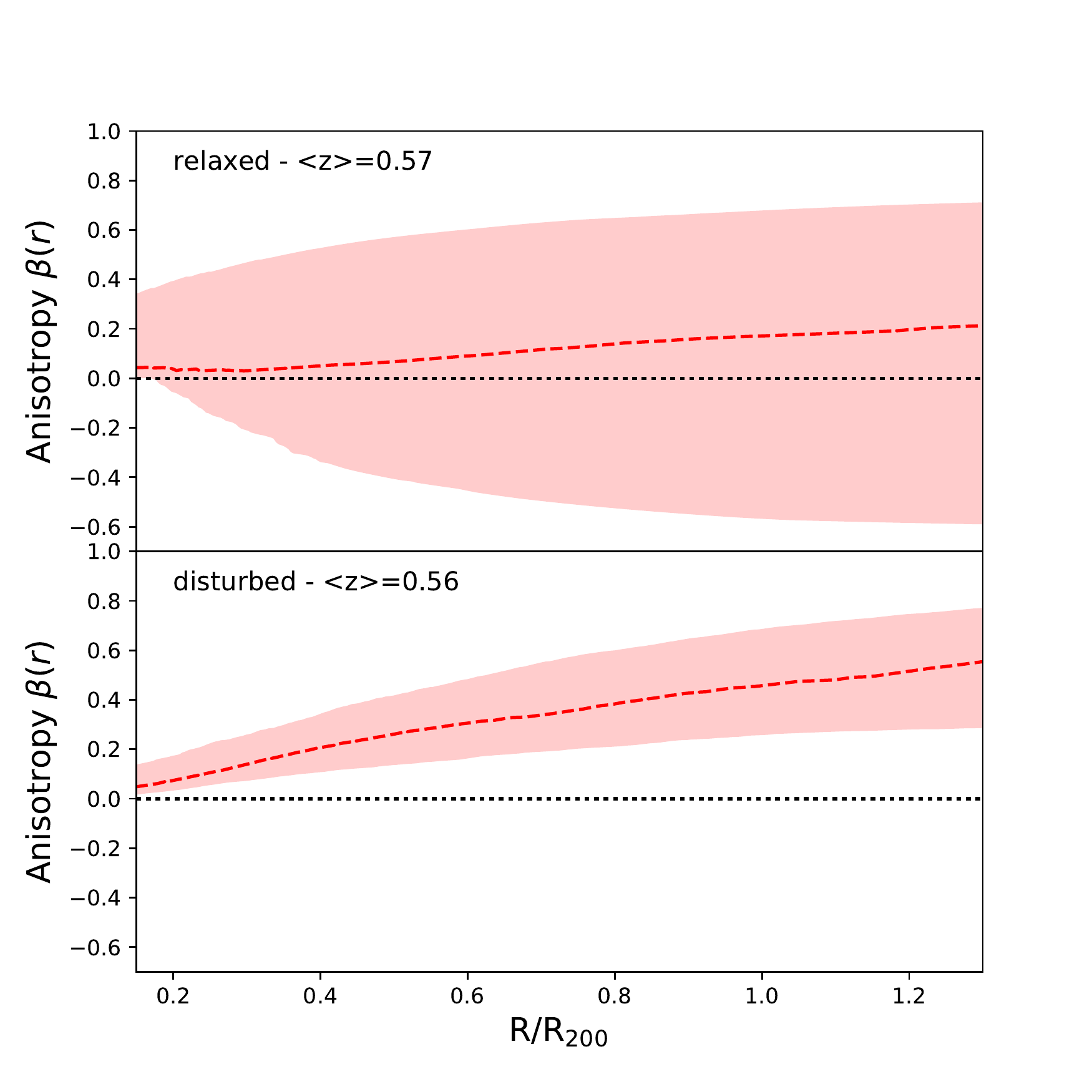}
\vskip-0.13in
\caption {Velocity anisotropy profile $\beta(r)$ recovered when separately analyzing relaxed and un-relaxed clusters. For the latter, the anisotropy profile indicates that the galaxies in the clusters are moving on nearly isotropic orbits near the center and increasingly radial orbits out toward the virial radius $R_{200}$.  The recovered anisotropy profile for the relaxed clusters is consistent with these results, but exhibits much larger uncertainties that allow for a much broader range of galaxy orbits. }
\label{fig:betaDist}
\vskip-0.20in
\end{figure}

One assumption in applying the Jeans equation to analyze our sample is that the galaxies we are analyzing are in approximate dynamical equilibrium. Thus, it would seem important to remove the obviously disturbed clusters--- those undergoing or having undergone recent, major mergers.  As discussed in \citet{mohr93}, the asymmetry and isophotal ellipticity of the X-ray surface brightness distribution provide information about the merger state of galaxy clusters, which is generally superior to the constraints possible from the galaxy distribution \citep{geller82} due to higher signal to noise. However, neither of these measures are sensitive to mergers along the line of sight, where the galaxy velocity distribution typically provides more information \citep{dressler88}.  

Merger signatures from X-ray cluster surface brightness distributions were extracted uniformly from a large X-ray flux limited sample of nearby clusters, indicating that over half of them exhibit statistically significant centroid variations and that the bulk of them are elliptical \citep{mohr95}. That study provided clear evidence that even massive clusters at low redshift are still undergoing continued accretion of subclusters.  This observation is in agreement with expectations from structure formation within our standard $\Lambda$CDM model.  We can use a similar approach to identify major mergers (not along the line of sight) in our cluster sample. \citet{2017Nurgaliev} quantified the X-ray morphology of a subsample of 90 SZE selected galaxy clusters using a measure of the photon asymmetry $A_{phot}$ \citep{2013Nurgaliev} closely related to the centroid variation \citep{mohr93,mohr95} and power ratios \citep{1995Buote, 1996Buote}.  

To test the dependence of our results on the dynamical state of the clusters, we adopt these $A_{phot}$ values as a measure of departures from equilibrium and separately analyze relaxed and disturbed clusters.  We take here relaxed clusters to be those with $A_{phot} \leq 0.2$.  Out of our 110 cluster sample, 68 (2152 spectra) have measured $A_{phot}$ and 39 of these (1258 galaxies) are classified as relaxed.   In Table~\ref{tab:results xrays} we show the results for the mass $M_{200}^{\text{dyn}}$ and concentration $c_{200}$ having performed the analysis on the stacked relaxed/disturbed populations.   We find that the \Mszsigma\ and the dynamical masses are in agreement at $\approx1.5\sigma$ level for the relaxed sample and at the $1\sigma$ level for the disturbed sample (statistical only).

Fig.~\ref{fig:betaDist} shows the recovered anisotropy profiles for the two subsamples. The anisotropy profile recovered for the disturbed clusters indicates that the galaxies are moving on roughly isotropic orbits in the center and increasingly radial orbits at large distance from the cluster core.  This is consistent with the behavior seen in the total sample and most subsamples. 
On the other hand, we find that the anisotropy profile of relaxed clusters, while consistent with this behavior, exhibits much larger uncertainties that allow also for very different behavior, including simple isotropic orbits at all radii.  A comparison shows that in the case of the relaxed sample one of the anisotropy models, namely the $\beta_{O}(r)$ one, which has anisotropy of opposite sign in the center and at large radii, has the highest Bayes factor.  In the case of this sample, the preference is for radial orbits near the center and somewhat tangential orbits at larger radius.  This is responsible for the extension of the uncertainties to anisotropies of $\sim-0.4$ at $R/R_{200}\sim1$.  We note that this kind of behavior is also shown in the upper right panel of Fig.~\ref{fig:anisotropy-redshift}, which is the redshift subsample that also prefers the $\beta_{O}$ model (see Table~\ref{tab:mamres}).

The differences in the character of these results is intriguing and deserves further study with larger dynamical samples in cluster ensembles that have associated measurements of substructure.  We note, however, that the results presented in this section on the velocity anisotropy profiles, the PPSD profiles and the cluster halo masses have been compared to studies of clusters formed in numerical structure formation simulations \citep[e.g.][]{1996NFW,2005Dehnen,2007Faltenbacher,2013MAMPOSSt} in which cluster substructure is generic.  Observationally, substructure has been established as generic to the real cluster population at low redshift for a timespan approaching four decades \citep{geller82}.  With a uniform selection of cluster subsamples defined in similar ways in simulations and the real world, it should be possible in future analyses to sharpen studies like ours to measure differences in galaxy orbital characteristics associated with X-ray substructure, presence of cool cores and so on.


\section{Conclusions}
\label{sec:conclusions}

We present a dynamical analysis of 110 SZE selected galaxy clusters from the SPT-SZ survey with redshifts between 0.2 to 1.3, that have an associated spectroscopic sample of more than 3000 passive galaxies.  We examine subsets of this cluster sample in redshift and mass, each comprising $\sim$600 cluster members.  These subsets are either combined to form composite clusters or are analyzed individually using a Jeans equation based code called MAMPOSSt \citep{2013MAMPOSSt} that allows one to adopt different parametric models for the mass profile, galaxy profile and velocity dispersion anisotropy profile.  In our analysis we adopt an NFW mass profile, and use the measured concentration of the red sequence galaxy population from a complete subsample of the SPT SZE selected cluster sample (H17), and employ five different velocity dispersion anisotropy profiles (see Section~\ref{sec:profiles}).  We perform Bayesian model averaging to combine results from the different dispersion anisotropy models, because none of the five models are excluded by the data.

The velocity dispersion anisotropy profiles show the same radial features at all redshifts: orbits are isotropic near the center and increasingly radial at larger radii. We also find no variations with cluster mass.  The radial variation is broadly consistent with that seen in a recent analysis of near-infrared selected clusters at $z\sim1$ \citep{2016Biviano} and also studies at low redshift \citep[$z\lesssim0.1$;][]{2009Lemze, 2010Wojtak}.  These trends of anisotropy with radius resemble those of DM particles in halos extracted from cosmological numerical simulations \citep[][and references therein]{2005Mamon, Mamon2010, 2013MAMPOSSt}.  The absence of a redshift trend is inconsistent with the results presented in \citet{2009Biviano}, where they report that passive galaxy orbits are becoming more isotropic over time.  
The absence of a redshift trend in the velocity anisotropy profiles suggests that the process of infall and relaxation for the passive galaxy population is occurring similarly at all redshifts since at least $z\sim1$.

We measure the pseudo-phase-space density (PPSD) profiles $Q_\mathrm{gal}(r)$ and $Q_\mathrm{gal,r}(r)$, using quantities derived from cluster galaxies. We find good agreement with theoretical predictions from $N$-body simulations of DM particles \citep{2005Dehnen}.  We examine whether the amplitude of the profile scales as expected with redshift and mass under the assumption of self-similarity, finding that they do.  To the extent that the PPSD profile provides constraints on the equilibrium nature of the galaxy dynamics, the good agreement with simulations suggests that galaxies behave approximately as collisionless particles and are as relaxed as the DM particles in halos forming within cosmological structure formation simulations.  Moreover, the lack of evidence for redshift trends in the power law index of the PPSD profiles suggests again that the passive galaxy population in clusters is dynamically similar at all redshifts and mass ranges probed in our study.

We carry out a consistency check between our dynamical masses $M_{200}^{\text{dyn}}$, which are marginalized over uncertainties in the velocity dispersion anisotropy profiles, with masses calibrated using the SPT cluster counts with and without strong external cosmological priors.  We find that our masses are smaller than those derived with strong external cosmological priors \MszPlanck\ by $\eta=0.81\pm0.06\pm0.08$, corresponding to a 1.9$\sigma$ discrepancy when systematic uncertainties in our dynamical masses are included.  Moreover, our analysis shows that the agreement is best at low redshift, while lower values of $\eta$ are preferred at higher redshift.  In the highest redshift bin the best fit mass ratio is $\eta=0.63^{+0.13}_{-0.08}\pm0.06$, which corresponds to a disagreement at the $2.6\sigma$ level. 

In addition, we find good agreement, $\eta'=1.14\pm0.07\pm0.11$, between our dynamical masses and those masses extracted from the SPT cluster counts in combination with 82 externally calibrated X-ray $Y_X$ mass estimates \citep{2016deHaan} when the cosmological parameters are allowed to vary \citep[and also those masses calibrated in combination with 63 velocity dispersions; see][]{bocquet15}. Our mass constraints are also consistent with those from related studies of SPT selected clusters, using both weak lensing magnification \citep{2016Chiu-a} and tangential shear \citep{dietrich17,2018Schrabback, stern18}.  

Using \textit{Chandra} X-ray data, we examine the impact of the dynamical state of the clusters on our dynamical analysis by separately analysing relaxed and un-relaxed clusters.  We find dynamical masses to be in good agreement with our combined sample for both the relaxed and un-relaxed clusters. Concerning the anisotropy profiles, we  find that, for the disturbed sample, the shape of the orbits resembles the one seen in the total sample and most subsamples. On the other hand, the anisotropy profile of relaxed clusters, while still consistent with this behavior, exhibits much larger uncertainties that allow also for isotropic orbits at all radii. Further investigation with larger dynamical samples in cluster ensembles is required in order to understand the different behaviour of these objects.

As a next step, our analysis can be extended to cluster samples that include many low mass systems.  One such sample that is being analyzed presently has been defined in the project known as SPIDERS \citep[SPectroscopic IDentification of eROSITA Sources, ][]{2016Clerc}, an optical spectroscopic survey of X-ray-selected galaxy clusters discovered in ROSAT and {\it XMM-Newton} imaging.   Another sample is being built up through spectroscopic observations of optically selected clusters within the Dark Energy Survey.  Longer term, we expect deep spectroscopic followup of SZE and X-ray selected clusters to provide ever larger galaxy samples that include both emission line and passive galaxies.  These samples will allow cluster masses to be constrained  in a redshift regime where weak lensing is challenging, while also enabling studies of the kinematic relationship between cluster emission line and passive galaxies out to redshifts well beyond 1.

\section*{ACKNOWLEDGMENTS}
We thank Crist\'obal Sif\'on for providing useful feedback. We acknowledge the support by the DFG Cluster of Excellence ``Origin and Structure of the Universe'', the Transregio program TR33 ``The Dark Universe'' and the Ludwig-Maximilians University. The South Pole Telescope is supported by the National Science Foundation through grant PLR-1248097. Partial support is also provided by the NSF Physics Frontier Center grant PHY-1125897 to the Kavli Institute of Cosmological Physics at the University of Chicago, the Kavli Foundation and the Gordon and Betty Moore Foundation grant GBMF 947. The Melbourne group acknowledges support from the Australian Research Council's Discovery Projects funding scheme (DP150103208). DR is supported by a NASA Postdoctoral Program Senior Fellowship at NASA's Ames Research Center, administered by the Universities Space Research Association under contract with NASA. Work at Argonne National Laboratory was supported under U.S. Department of Energy contract DE-AC02-06CH11357. AB acknowledges the hospitality of the LMU, and partial financial support from PRIN-INAF 2014 "Glittering kaleidoscopes in the sky: the multifaceted nature and role of Galaxy Clusters?, P.I.: Mario Nonino. BB has been supported by the Fermi Research Alliance, LLC under Contract No. DE-AC02-07CH11359 with the U.S. Department of Energy, Office of Science, Office of High Energy Physics. AS is supported by the ERC-StG ?ClustersXCosmo", grant agreement 716762.

\bibliographystyle{mn2e}
\bibliography{paper,spt}

\section*{Affiliations}

\textit{
\Munich Faculty of Physics, Ludwig-Maximilians-Universit\"{a}t, Scheinerstr.\ 1, 81679 Munich, Germany \\
\ExcellenceCluster Excellence Cluster Universe, Boltzmannstr.\ 2, 85748 Garching, Germany \\
\Trieste  INAF-Osservatorio Astronomico di Trieste via G.B. Tiepolo 11, 34143 Trieste, Italy \\
\MPE Max Planck Institute for Extraterrestrial Physics, Giessenbachstr.\ 85748 Garching, Germany \\
\KICPChicago Kavli Institute for Cosmological Physics, University of Chicago, 5640 South Ellis Avenue, Chicago, IL 60637 \\
\MIT Kavli Institute for Astrophysics and Space Research, Massachusetts Institute of Technology, 77 Massachusetts Avenue, Cambridge, MA 02139, USA \\
\AAUChicago Department of Astronomy and Astrophysics, University of Chicago, Chicago, IL, USA 60637 \\
\FNAL Fermi National Accelerator Laboratory, Batavia, IL 60510-0500, USA \\
\PhysicsUChicago Department of Physics, University of Chicago, Chicago, IL, USA 60637 \\
\ANL Argonne National Laboratory, 9700 S. Cass Avenue, Argonne, IL, USA 60439 \\
\Miss Department of Physics and Astronomy, University of Missouri, 5110 Rockhill Road, Kansas City, MO 64110 \\
\EFIChicago Enrico Fermi Institute, University of Chicago, Chicago, IL, USA 60637 \\
\Taipei Institute of Astronomy and Astrophysics, Academia Sinica, Taipei 10617, Taiwan \\
\Berkeley Department of Physics, University of California, Berkeley, CA, USA 94720 \\
\McGill Department of Physics, McGill University, Montreal, Quebec H3A 2T8, Canada \\
\StanfordKPAC Kavli Institute for Particle Astrophysics and Cosmology, Stanford University, 452 Lomita Mall, Stanford, CA 94305-4085, USA \\
\StanfordPhys Department of Physics, Stanford University, 382 Via Pueblo Mall, Stanford, CA 94305 \\
\UMon Department of Physics, Universit\'e de Montr\'eal, Montrea, Quebec H3T 1J4, Canada \\
\StonyBrook Department of Physics and Astronomy, Stony Brook University, Stony Brook, NY 11794, USA \\
\DARK Dark Cosmology Centre, Niels Bohr Institute, University of Copenhagen Juliane Maries Vej 30, 2100 Copenhagen, Denmark \\
\Colorado Center for Astrophysics and Space Astronomy, Department of Astrophysical and Planetary Sciences, University of Colorado, Boulder, CO, 80309\\
\NASA NASA Ames Research Center, Moffett Field, CA 94035, USA \\
\UM School of Physics, University of Melbourne, Parkville, VIC 3010, Australia \\
\Michigan Department of Astronomy, University of Michigan, 1085 S. University Ave, Ann Arbor, MI 48109, USA \\
\CfA Harvard-Smithsonian Center for Astrophysics, Cambridge, MA, USA 02138 \\
\Davis Department of Physics, University of California, Davis, CA, USA 95616 \\
\LLNL Institute of Geophysics and Planetary Physics, Lawrence Livermore National Laboratory, Livermore, CA 94551 \\
\Chile Cerro Tololo Inter-American Observatory, Casilla 603, La Serena, Chile \\
}

\end{document}